\documentclass[11pt]{amsart}
\reversemarginpar
\newcommand{\commentout}[1]{}

\usepackage{amsmath}
\usepackage{amssymb}

\newcommand{\nwc}{\newcommand}

\newcommand{\partz}{\frac{\partial }{\partial z}}
\newcommand{\invf}{\cF^{-1}_2}

\newcommand{\al}{\alpha}

\newcommand{\kvec}{{\vec{\mathbf k}}}
\newcommand{\lt}{\left}
\newcommand{\rt}{\right}
\newcommand{\vas}{\varepsilon}
\newcommand{\lan}{\left\langle}
\newcommand{\ran}{\right\rangle}
\newcommand{\tvas}{W^\ep_z}
\newcommand{\psiep}{W^\ep_z}

\newcommand{\wepz}{{W^\ep_z}}

\newcommand{\cv}{{\ml L}_z}
\newcommand{\cvtil}{\tilde{\ml L}^{*}_z}

\newcommand{\ks}{{k}}
\newcommand{\bx}{\mb x}

\newcommand{\bD}{\mathbf D}
\newcommand{\bp}{\mathbf p}
\newcommand{\bq}{\mathbf q}
\newcommand{\by}{\mathbf y}

\newcommand{\pdgx}{\bp\cdot\nabla_\bx}

\nwc{\xtil}{\tilde{\bx}}
\nwc{\epal}{\ep^{-2\al}}

\newcommand{\cltil}{\tilde{\cL}^{*}}
\newcommand{\fW}{\mathfrak{W}}

\nwc{\nwt}{\newtheorem}

\nwt{prop}{Proposition}
\nwt{proposition}{Proposition}
\nwt{lemma}{Lemma}
\nwt{assumption}{Assumption}
\nwt{theorem}{Theorem}
\nwt{cor}{Corollary}
\nwt{corollary}{Corollary}
\nwt{remark}{Remark}
\nwt{definition}{Definition} 

\nwc{\bal}{\begin{align}}
\nwc{\be}{\begin{equation}}
\nwc{\ben}{\begin{equation*}}
\nwc{\bea}{\begin{eqnarray}}
\nwc{\beq}{\begin{eqnarray}}
\nwc{\bean}{\begin{eqnarray*}}
\nwc{\beqn}{\begin{eqnarray*}}
\nwc{\beqast}{\begin{eqnarray*}}

\nwc{\eal}{\end{align}}
\nwc{\ee}{\end{equation}}
\nwc{\een}{\end{equation*}}
\nwc{\eea}{\end{eqnarray}}
\nwc{\eeq}{\end{eqnarray}}
\nwc{\eean}{\end{eqnarray*}}
\nwc{\eeqn}{\end{eqnarray*}}
\nwc{\eeqast}{\end{eqnarray*}}

\nwc{\ep}{\varepsilon}
\nwc{\eps}{\varepsilon}
\nwc{\ept}{\epsilon}
\nwc{\vrho}{\varrho}
\nwc{\orho}{\bar\varrho}
\nwc{\ou}{\bar u}
\nwc{\vpsi}{\varpsi}
\nwc{\lamb}{\lambda}

\nwc{\nn}{\nonumber}
\nwc{\bm}{\boldmath}
\nwc{\mf}{\mathbf}
\nwc{\mb}{\mathbf}
\nwc{\ml}{\mathcal}

\nwc{\IA}{\mathbb{A}} 
\nwc{\IB}{\mathbb{B}}
\nwc{\IC}{\mathbb{C}} 
\nwc{\ID}{\mathbb{D}} 
\nwc{\IM}{\mathbb{M}} 
\nwc{\IP}{\mathbb{P}} 
\nwc{\II}{\mathbb{I}} 
\nwc{\IE}{\mathbb{E}} 
\nwc{\IF}{\mathbb{F}} 
\nwc{\IG}{\mathbb{G}} 
\nwc{\IN}{\mathbb{N}} 
\nwc{\IQ}{\mathbb{Q}} 
\nwc{\IR}{\mathbb{R}} 
\nwc{\IT}{\mathbb{T}} 
\nwc{\IZ}{\mathbb{Z}} 

\nwc{\cE}{{\ml E}}
\nwc{\cP}{{\ml P}}
\nwc{\cQ}{{\ml Q}}
\nwc{\cL}{{\ml L}}
\nwc{\cX}{{\ml X}}
\nwc{\cW}{{\ml W}}
\nwc{\cZ}{{\ml Z}}
\nwc{\cR}{{\ml R}}
\nwc{\cV}{{\ml L}}
\nwc{\cT}{{\ml T}}
\nwc{\crV}{{\ml L}_{(\delta,\rho)}}
\nwc{\cC}{{\ml C}}
\nwc{\cA}{{\ml A}}
\nwc{\cK}{{\ml K}}
\nwc{\cB}{{\ml B}}
\nwc{\cD}{{\ml D}}
\nwc{\cF}{{\ml F}}
\nwc{\cS}{{\ml S}}
\nwc{\cM}{{\ml M}}
\nwc{\cG}{{\ml G}}
\nwc{\cH}{{\ml H}}
\nwc{\bk}{{\mb k}}
\nwc{\cbz}{\overline{\cB}_z}

\newcommand{\grx}{\nabla_\bx}
\newcommand{\gry}{\nabla_\by}

\newcommand{\grxtil}{\nabla_{\tilde\bx}}

\nwc{\pft}{\cF^{-1}_2}

\begin{document}

\title{Self-Averaging Scaling Limits of Two-Frequency
Wigner Distribution for Random Paraxial Waves}

\author{Albert C. Fannjiang
 }
\thanks{Department of Mathematics,
University of California,
Davis, CA 95616-8633
Email: fannjiang@math.ucdavis.edu.
The research is supported in part by National Science Foundation grant no. DMS-0306659, ONR Grant N00014-02-1-0090
and  Darpa Grant 
 N00014-02-1-0603
}

\begin{abstract} Two-frequency
Wigner distribution is introduced to capture the asymptotic 
behavior of the space-frequency correlation of paraxial waves in the radiative
transfer limits. The scaling limits give rises to 
deterministic transport-like equations. 
Depending on the ratio of the wavelength to the correlation length
the limiting equation is either a Boltzmann-like 
integral equation or a Fokker-Planck-like differential
equation in the phase space. The solutions to these
equations have a probabilistic representation which
can be simulated by Monte Carlo method. 
When the medium fluctuates more rapidly in the longitudinal direction, the corresponding  Fokker-Planck-like equation can be solved
exactly. 
\end{abstract}

\maketitle

\section{Introduction}
A central quantity of wave propagation in random media
is the correlation of wave field at two space-time points.
Through
spectral decomposition of the time-dependent
signal the space-time correlation is equivalent to
the space-frequency correlation of the wave field \cite{MW}.
The main focus of the present work is
to derive rigorously closed form equations governing
the space-frequency correlation of  paraxial waves
in the radiative transfer regime.

For 
optical wave propagating through the turbulent atmosphere,
the complex-valued  wave amplitude  is governed by the stochastic Schr\"odinger (paraxial) equation 
 with a white-noise potential
\cite{St}. The conventional approach uses  
the two-frequency mutual coherence
function and  various ad hoc approximations \cite{Ish}.  
 Recently,  we introduced the two-frequency
Wigner distribution in terms of which we derived
rigorously a complete
set of two-frequency all-order moment equations and 
solved exactly the mutual coherence 
function  in the geometrical optics
regime \cite{2f-whn}.

In this paper, we consider the different
regime of  radiative transfer
and prove the {\em self-averaging} convergence of
the two-frequency Wigner distribution for
the paraxial wave equation. In other words,
in radiative transfer, the whole hierarchy of
two-frequency moment equations is reduced to a single
radiative-transfer-like equation. 

Let $L_z$ and $L_x$ be, respectively, the characteristic
  macroscopic length scales of the wave beam in the longitudinal and transverse
  directions. We assume the phase speed in the vacuum is unity. The Fresnel number  commonly defined
  as
  \[
F=  \frac{L_x^2}{\lambda_0 L_z}
  \] 
with the central wavelength $\lambda_0=2\pi/k_0$
measures the significance of Fresnel diffraction. In all
the scalings considered here the corresponding Fresnel number is large,
indicating strong Fresnel diffraction effect. 
  
We use $\lambda_0, L_z, L_x$ to non-dimensionalize the paraxial wave
equation \cite{Ish}. Let $k_1, k_2$ be two (relative) wavenumbers nondimensionalized by the central wavenumber $k_0$.  Then
 the wave field $\Psi_j$ of
wavenumber  $k_j$
satisfies 
\beq
\label{para2}
i\partz
\Psi_j(z,\bx)+\frac{\gamma}{2\ks_j}\nabla_\bx^2\Psi_j(z,\bx)
+
\frac{\mu k_j}{\gamma}V(\frac{zL_z}{\ell_z}, \frac{\bx L_x}{\ell_x})\Psi_j(z,\bx)=0, \quad j=1,2
\eeq
where  $\gamma$ is the reciprocal of  Fresnel number,
$V$ represents 
 the refractive index  fluctuation with
  the correlation lengths $\ell_z$ and $\ell_x$
  in the longitudinal and transverse directions,
  respectively, and $\mu$ is the magnitude of
  the fluctuation. We denote the ratios $\ell_z/L_z$ and
  $\ell_x/L_x$, respectively, by $\rho_z$ and $\rho_x$
  which then are the correlation lengths of the medium fluctuations in the unit of
  characteristic   length scales of the wave beam. 

Let us now describe a family of scaling limits which would yield
non-trivial
self-averaging limit for the space-frequency correlation 
of the wave beam. 
The main characteristic of the scaling limits is that the transverse correlation length  
$\ell_x$ is much smaller than
the beam width $ L_x$ but is comparable or
much larger than the central wavelength $\lambda_0$.
Roughly speaking, the wave beam experiences
the spatial diversity as it propagates through the
medium but it 
does not do so over a single wavelength
in order to avoid  the effective medium regime.
This condition is best described by dimensionless quantities  as
\beq
\label{sc1}
\theta\equiv \gamma/\rho_x=O(1)
\eeq
which includes the possibility that $\theta\ll 1$. 
This is the 
radiative transfer regime  for monochromatic waves 
\cite{rad-crm}. 

To reduce the complexity of technique, however,  we restrict ourselves
mainly to the longitudinal case where $\rho_z/\rho_x=O(1)$, including
the possible scenario $\rho_z/\rho_x\ll 1$. 
In this case,  the medium fluctuation
of the following order of magnitude 
\bea
\label{sc2}
{\mu}=\frac{\gamma}{\theta \sqrt{\rho_z}}=\frac{\rho_x}{\sqrt{\rho_z}}
\eea
 would lead to
non-trivial multiple scattering effect
(see \cite{rad-arma}, \cite{rad-crm} and references therein).

For wave fields of  two different frequencies to interfere coherently, 
the  unscaled frequency shift needs to be limited 
to $O(1)$ regardless the central frequency. In other words, we look for $O(1)$ coherence bandwidth which is a  characteristic of the random medium. Again,
this is best expressed in the dimensionless quantities as
\beq
\label{band}
\label{sc3}
\lim_{\ell_x\to 0}\ks_1=\lim_{\ell_x\to 0}\ks_2=\ks,
\quad \lim_{\ell_x\to 0}\gamma^{-1}k^{-1}(\ks_2-\ks_1)=\beta. 
\eeq
We assume $\beta>0$ below. We call this  the two-frequency radiative transfer  scaling limits.

Let us now review the basic framework
of two-frequency Wigner distribution in the paraxial regime.
The two-frequency Wigner distribution is defined as
\cite{2f-whn}
\beq
\label{0.11}
W(z,\bx,\bp)=\frac{1}{(2\pi)^d}\int
e^{-i\bp\cdot\by}
\Psi_1 (z,\frac{\bx}{\sqrt{\ks_1}}+\frac
{\gamma\by}{2\sqrt{\ks_1}}){\Psi_2^{*}(z,\frac{\bx}{
\sqrt{\ks_2}}
-\frac{\gamma\by}{2\sqrt{\ks_2}})}d\by
\eeq
where the scaling factor $\sqrt{k_j}$ is introduced 
so that $W$ satisfies a closed-form equation (see below). Note
here the scaling factor for the parabolic wave is
different from that for the spherical wave introduced
in \cite{2f-rt-josa}.

The following property can be derived easily from
the definition 
\beqn
\|W(z,\cdot,\cdot)\|_2&=&
\lt(\frac{\sqrt{\ks_1\ks_2}}{2\gamma\pi}\rt)^{d/2}
\|\Psi_1(z,\cdot)\|_2\|\Psi_2(z,\cdot)\|_2.
\eeqn
Since $\Psi_j$ are governed by the Schr\"odinger-like equation  the $L^2$-norm does not change with $z,$ i.e.,  $
\|W(z,\cdot,\cdot)\|_2=
\|W(0,\cdot,\cdot)\|_2
$.
The Wigner distribution 
has the following obvious properties:
\bea\label{2.2.2}
\int W(z,\bx,\bp)e^{i\bp\cdot\by}d\bp&=&
\Psi_1
(z,\frac{\bx}{\sqrt{\ks_1}}+
\frac{\gamma\by}{2\sqrt{\ks_1}})
\Psi_2^{*}(z,\frac{\bx}{
\sqrt{\ks_2}}
-\frac{\gamma\by}{2\sqrt{\ks_2}})\\
\int_{\IR^d}W(z,\bx,\bp)e^{-i\bx\cdot
\bq}d\bx&=&\lt(\frac{\pi^2\sqrt{\ks_1\ks_2}}{\gamma}\rt)^{d}
\widehat\Psi_1(z,\frac{\bp\sqrt{\ks_1}}{4\gamma}
+\frac{\sqrt{\ks_1}\bq}{2})
{\widehat\Psi}^{*}_2(z,\frac{\bp\sqrt{\ks_2}}{4\gamma}
-\frac{\sqrt{\ks_2}\bq}{2})
\eeq
and so contains essentially all the information in the
two-point two-frequency function. 

The Wigner distribution $W_z$
satisfies the Wigner-Moyal equation 
exactly \cite{2f-whn}
\beq
\frac{\partial W}{\partial z}
+{\bp}\cdot\nabla_\bx W
+\frac{1}{\sqrt{\rho_z}}\cv W=0
\label{wig}
\eeq
\commentout{
with the initial data
\beq
\label{pure}
W_0(\bx,\bk)=\frac{1}{(2\pi)^d}\int e^{i\bk\cdot\by}
\Psi_{1,0}
(\frac{\bx}{\sqrt{\ks_1}}+\frac{\gamma\by}{2\sqrt{\ks_1}}
){\Psi^*_{2,0}}(\frac{\bx}{
\sqrt{\ks_2}}
-\frac{\gamma\by}{2\sqrt{\ks_2}})d\by \,,
\eeq
}
where the operator $\cv$ is formally given as
\beq
\label{L}
\cv W&=&
i\int 
\theta^{-1}\lt[e^{i\bq\cdot\xtil/\sqrt{\ks_1}}\ks_1
W(z, \bx,\bp+\frac{\theta\bq}{2\sqrt{\ks_1}})-
e^{i\bq\cdot\xtil/\sqrt{\ks_2}}\ks_2W(z,\bx,\bp-
\frac{\theta\bq}{2\sqrt{\ks_2}})\rt]
\widehat{V}(\frac{ z}{\rho_z},d\bq)
\nn
\eeq
with $\xtil=\bx/\rho_x$ being the `fast' transverse variable. 
As $\rho_z\to 0, \rho_z^{-1/2}\cL_z$ displays 
the classical central limit scaling in the $z$-variable and
thus this is referred to as the longitudinal case. 
The complex conjugate ${\wepz}^{*}(\bx,\bp)$
satisfies a similar  equation
\beq
\frac{\partial {W}^*}{\partial z}
+{\bp}\cdot\nabla_\bx {W}^*
+\frac{1}{\sqrt{\rho_z}}\cv^* W^*=0
\label{wigcc}
\eeq
where
\beq
\label{Lcc}
\cv^*W^*&=&
i\int 
\theta^{-1}\lt[e^{i\bq\cdot\xtil/\sqrt{\ks_2}}\ks_2
W^*(z,\bx,\bp+\frac{\theta\bq}{2\sqrt{\ks_2}})-
e^{i\bq\cdot\xtil/\sqrt{\ks_1}}\ks_1W^*(z,\bx,\bp-
\frac{\theta\bq}{2\sqrt{\ks_1}})\rt]
\widehat{V}(\frac{z}{\rho_z},d\bq).
\nn
\eeq

We use
the following definition of the Fourier transform 
and inversion:
\beqn
\cF f(\bp)&=&\frac{1}{(2\pi)^d}\int e^{-i\bx\cdot\bp}
f(\bx)d\bx\\
\cF^{-1} g(\bx)&=&\int e^{i\bp\cdot\bx} g(\bp) d\bp.
\eeqn
When making a {\em partial}  (inverse) Fourier transform on
a phase-space function we will write $\cF_1$ (resp. $\cF^{-1}_1$) 
and
$\cF_2$ (resp. $\cF^{-1}_2$) 
to denote the  (resp. inverse) transform w.r.t. $\bx$ and $\bp$
respectively.

\section{Formulation and theorems} 
To describe the scaling limits more efficiently we introduce
two  controlling parameters $\ep, \alpha>0$ and set 
\beq
\label{sc4}
\rho_z=\ep^{2},\quad\rho_x=\ep^{2\alpha}. 
\eeq
By (\ref{sc1})-(\ref{sc3}) all the other parameters can be
expressed in terms of $\ep$, $\alpha$ and $\theta$. 
The radiative transfer scalings in the longitudinal case then correspond to
\beq
\label{rtsc}
\ep\to 0,\quad \alpha\in (0,1],\quad \lim_{\ep\to 0}\theta<\infty
\eeq
along with (\ref{sc1})-(\ref{sc3}).

   
To get the self-averaging result, it is essential to consider the
weak formulation  of the Wigner-Moyal  equation:
To find $W^\ep_z$ in the space  $C([0,\infty);
L^2_w(\IR^{2d})$ of $z$-continuous, 
$L^2(\IR^{2d})$-valued processes 
such that
$\|W^\ep_z\|_2\leq \|W_0\|_2, \forall z>0,$ and 
\beq
\lan W^\ep_z, \Theta\ran - \lan W_0,
\Theta \ran &=&
\int_0^z \lan W^\ep_s, \pdgx \Theta\ran ds
+\frac{1}{\vas}\int_0^z \lan W^\ep_s, {\cL}^{*}_s
\Theta
\ran ds,
\label{weak}
\eeq
where $W_0\in L^2(\IR^{2d})$ is the initial condition and  we consider all the smooth test functions of compact
supports 
$ \Theta\in C_c^\infty(\IR^{2d})$.
 Here and below $L^2_w(\IR^{2d})$ is the space of
complex-valued square integrable
functions on the phase space $\IR^{2d}$ endowed with
the weak topology and the inner product
\[
\lan W_1, W_2\ran =
\int W_1^*(\bx,\bp)W_2(\bx,\bp)d\bx d\bp.
\]

 We define for every realization
of $V^\ep_z$ the operator $\cv^*$ to act on a 
phase-space test function
$\Theta$ as
\beq
\cv^*\Theta(\bx,\bp)
\equiv  -i\theta^{-1}\cF_2\lt[\delta_{\ep}
V^\ep_z(\bx,\by) \cF^{-1}_2
\Theta(\bx,\by)\rt]
\label{cv}
\eeq
with the difference operator $\delta_{\ep}$ given by 
\beq
\label{diff}
\delta_{\ep} V^\ep_z(\bx,\by)&\equiv&
 \ks_1
V^\ep_z(\frac{\bx}{\sqrt{\ks_1}}+
\frac{\theta\by}{2\sqrt{\ks_1}})
-
k_2V^\ep_z(\frac{\bx}{\sqrt{\ks_2}}-
\frac{\theta\by}{2\sqrt{\ks_2}})
\eeq
We define $\cv$ in the similar way.

The existence of solutions in the space  $C([0,\infty);
L^2_w(\IR^{2d}))$ can
be established by  the same weak compactness
argument as in \cite{2f-whn}. We will
not, however,  address the uniqueness of solution for the Wigner-Moyal
equation (\ref{weak}) but we will show that  as $\ep\to 0$
any sequence of weak solutions to eq. (\ref{weak}) converges
in a suitable sense to the unique solution of a deterministic equation (see Theorem~1 and 2).

We assume that $V_z(\bx)=V(z,\bx)$ is a real-valued, centered,  $z$-stationary,
$\bx$-homogeneous ergodic random field
admitting the spectral representation
\[
V_z(\bx)=\int \exp{(i\bp\cdot\bx)}\hat{V}_z(d
\bp)
\]
with the  $z$-stationary spectral measure
$\hat{V}_z(\cdot)$ satisfying
\[
\IE[\hat{V}_z(d\bp)\hat{V}_z(d\bq)]
=\delta(\bp+\bq)\Phi_0(\bp)d\bp d\bq.
\]

The transverse power spectrum  density is
related to the full power spectrum density
$\Phi(w,\bp)$ as $\Phi_0(\bp)=\int \Phi(w,\bp)dw.
$
The power spectral density $\Phi(\kvec)$
satisfies $\Phi(\kvec)=\Phi(-\kvec),\forall
 \kvec=(w, \bp)\in \IR^{d+1}$
because
 the electric susceptibility field is assumed to be real-valued.
Hence
$
\Phi(w,\bp)=\Phi(-w,\bp)=\Phi(w,-\bp)
=\Phi(-w,-\bp)
$
which is related to the detailed balance of
the limiting scattering operators
described below.

The first main assumption is the Gaussian property
of the random potential. 
\begin{assumption}
 $V(z, \bx) $ is a   Gaussian process  with a
spectral density $\Phi(\kvec), \kvec=(w,\bp)\in\IR^{d+1}$ which is smooth, uniformly bounded
and decays at $|\kvec|=\infty$ with sufficiently
high power of $|\kvec|^{-1}$.
\end{assumption}
We note that the assumption of Gaussianity is not
essential and is made here to simplify the presentation.
Its main use is in deriving the estimates (\ref{G1}), (\ref{G2})
below.

Let $\cF_z $ and $\cF^+_z$  be the sigma-algebras generated by
$\{V_s:  \forall s\leq z\}$ and $\{V_s: \forall s\geq  z\}$,
respectively and let $L^2(\cF_z)$ and $L^2(\cF^+_z)$
denote the square-integrable functions measurable w.r.t.
to them respectively.
The maximal  correlation coefficient $\rho(t)$ is given by
\beq
\label{correl}
\rho(t)=\sup_{h\in L^2(\cF_z)\atop \IE[h]=0,
\IE[h^2]=1}
\sup_{
g\in L^2(\cF_{z+t}^+)\atop \IE[g]=0,
\IE[g^2]=1}\IE\lt[h g\rt].
\eeq

For  Gaussian processes
the correlation coefficient
$\rho(t)$ equals the linear correlation
coefficient  given by
\beq
\label{corr}
\sup_{g_1, g_2}
 \int R(t-\tau_1-\tau_2,\bk) g_1(\tau_1,\bk)
g_2(\tau_2,\bk)d\bk d\tau_1 d\tau_2
\eeq
where
$
R(t,\bk)=\int e^{it\xi} \Phi(\xi, \bk) d\xi
$
and the
supremum is taken over all $g_1, g_2\in L^2(\IR^{d+1})$ which are
supported on $(-\infty, 0]\times
\IR^d$ and  satisfy the constraint
\beqn
\int R(t-t',\bk) g_1(t,\bk){g}^*_1(t',\bk) dt dt' d\bk=
\int R(t-t',\bk) g_2(t,\bk){g}^*_2(t',\bk) dt dt' d\bk=1.
\eeqn
\commentout{
Alternatively, by the Paley-Wiener theorem we can write
\beq
\label{corr2}
L(t) &=&\sup_{f_1, f_2}
 \int e^{i\xi t}f_1(\xi,\bk)f_2(\xi,\bk)
\Phi(\xi,\bk)d\xi d\bk
\eeq
where $f_1, f_2$ are elements of the  Hardy space $\cH^2$
of $L^2(
\Phi(w,\bk) dw d\bk)$-valued analytic
functions in the upper half
$\xi$-space satisfying the normalization condition
\[
\int |f_j(\xi,\bk)|^2\Phi(\xi,\bk)d\xi d\bk=1, \quad j=1,2.
\]
}
There are various criteria for the decay rate of the linear correlation coefficients, see \cite{IR}.
 
Next we make assumption on the mixing property of the random potential. 
\begin{assumption}
\label{ass1}

The maximal correlation coefficient
$\rho(t)$  is integrable:
$\int^\infty_0\rho(s)ds
<\infty.
$
\end{assumption}

We have two main theorems depending on whether
 $\lim_{\ep\to 0}\theta$ is positive or not.

\begin{theorem}
Let $\theta>0$ be fixed.    Then  under
the two-frequency radiative transfer scaling (\ref{sc1})-(\ref{sc3}), (\ref{sc4}) the 
weak solutions, denoted by $W^\ep_z$,  of the Wigner-Moyal equation 
(\ref{wig})
converges in probability in 
$C([0,\infty), L^2_w(\IR^d))$, the space of
$L^2$-valued $z$-continuous processes,   to that of
the following deterministic equation
\beq
\label{bolt}
{\partz W_z+\bp\cdot\nabla  W_z}
&=&
\frac{2\pi k^2}{\theta^2}\int
K(\bp,\bq)
\lt[
e^{-i\beta\theta\bq\cdot\bx/(2\sqrt{k})}
W_z(\bx,\bp+\frac{\theta \bq}{\sqrt{k}})
-W_z(\bx,\bp)
\rt]
{d\bq}
\eeq
where the kernel $K$ is given by
\beqn
K(\bp,\bq)&=&
\Phi(0,\bq),\quad\hbox{for}\quad \alpha\in (0,1),
\nn
\eeqn
and 
\beqn
K(\bp,\bq)&=&\Phi\big((\bp+\frac{\theta\bq}{2\sqrt{k}})\cdot\bq,\bq\big),\quad
\hbox{for}\quad \alpha=1.
\nn
\eeqn

\end{theorem}
\begin{theorem}
Assume $\lim_{\ep\to 0} \theta=0$.  Then under the two-frequency radiative transfer
scaling (\ref{sc1})-(\ref{sc3}), (\ref{sc4})  the 
weak solutions of the Wigner-Moyal equation 
(\ref{wig})
converges in probability in the space
$C([0,\infty),L^2_w(\IR^d))$ to that of the following
deterministic equation
\beq
\label{fp}
\partz W_z+\bp\cdot\nabla W_z=\ks\lt(\nabla_\bp-\frac{i}{2}{\beta}\bx\rt)
\cdot \bD\cdot 
\lt(\nabla_\bp-\frac{i}{2}\beta\bx\rt) W_z
\eeq
where the (momentum) diffusion coefficient $\bD$ is
given by
\beq
\label{D1}\bD&=&{\pi}\int \Phi(0,\bq)\bq\otimes\bq d\bq, \quad\hbox{for}\quad \alpha\in (0,1),\\
\label{D3}
\bD(\bp)&=&{\pi}\int \Phi({\bp\cdot\bq},\bq)
\bq\otimes\bq d\bq, \quad\hbox{for}\quad \alpha=1.
\eeq
\end{theorem}

\begin{remark}
In the {\em transverse} case of $\rho_x\ll \rho_z$ (or $\alpha>1$), then with
the choice of $\mu=\sqrt{\rho_x}$ (or $\ep^\alpha$) the limiting kernel and diffusion coefficient
become, respectively
\beq
K(\bp,\bq)&=&
\delta\big((\bp+\frac{\theta\bq}{2\sqrt{k}})\cdot\bq\big)
\int\Phi(w,\bq)
dw,\\
\nn
\label{D2}
\bD(\bp)&=&\pi |\bp|^{-1}
\int_{\bp\cdot\bp_\perp=0} \int
\Phi(w,\bp_\perp) d w\,\,
\bp_\perp\otimes\bp_\perp
d\bp_\perp.
\eeq
The proof  of such result requires additional assumptions
which would complicate our presentation, so
we will not pursue it here, cf. \cite{rad-crm}, \cite{rad-arma}.
\end{remark}

\begin{remark}
The form of (\ref{bolt}) and (\ref{fp}) suggests the new
quantity 
\[
\fW_z(\bx,\bp)=e^{-\frac{i\beta}{2}\bx\cdot\bp}W_z(\bx,\bp)
\]
in terms of which the equations can be written respectively as
\beq
\label{bolt2}
{\partz \fW_z+\bp\cdot\nabla  \fW_z}+\frac{i\beta}{2}|\bp|^2\fW_z
&=&
\frac{2\pi k^2}{\theta^2}\int
K(\bp,\bq)
\lt[
\fW_z(\bx,\bp+\frac{\theta \bq}{\sqrt{k}})
-\fW_z(\bx,\bp)
\rt]
{d\bq},\\
\label{fp2}
\partz \fW_z+\bp\cdot\nabla \fW_z+\frac{i\beta}{2}|\bp|^2\fW_z&=&\ks\nabla_\bp
\cdot \bD\cdot
\nabla_\bp\fW_z.
\eeq
The advantage of this is that the solution then is amenable to
probabilistic representation as the scattering terms on the right side
of equations are non-positive definite. Let $(\bx(z),\bp(z))$ be
the stochastic process with $\bx(0)=\bx,\bp(0)=\bp$ generated by the operator
$-\bp\cdot\nabla_\bx+\cA$ where $\cA$ is the operator 
on the right hand side of either (\ref{bolt2}) or (\ref{fp2}). 
Then by Dynkin's formula we have
\beq
\label{MC}
\fW_z(\bx,\bp)=
\IE_{\bx,\bp}\lt[ e^{-\frac{i\beta}{2}\int^z_0 |\bp(t)|^2 dt}
e^{-\frac{i\beta}{2}\bx(z)\cdot\bp(z)}W_0(\bx(z),\bp(z))\rt]
\eeq
where $W_0$ is the initial data for the two-frequency Wigner distribution and
$\IE_{\bx,\bp}$ is the expectation with respect to
the probability measure associated with $(\bx(z),\bp(z))$.
The probabilistic solution (\ref{MC}) can be simulated
by Monte Carlo method. 
\end{remark}
When $k_1=k_2$ or $\beta=0$, eq. (\ref{bolt}) and (\ref{fp})
reduce to the standard radiative transfer equations derived
in \cite{rad-arma}, \cite{rad-crm}. 
In view of the definition (\ref{0.11}) the two-frequency Wigner distribution,
captures the space-frequency correlation on the nondimensionalized scale of $\gamma$
(in space as well as in frequency). The self-averaging property simply reflects
the fact that on the wave field essentially decorrelates
on the macroscopic scale $\gg \gamma$ and the rapidly fluctuating limit is
then statistically stablized by coupling with a smooth deterministic  test function.

A notable fact is that 
eq. (\ref{fp}) with (\ref{D1})
is the same governing equation, except for
a constant damping term,  for
the {\em ensemble-averaged }
two-frequency Wigner distribution for
the $z$-white-noise potential in
the geometrical optics regime, \cite{2f-whn}. 
Fortunately
eq. (\ref{fp})-(\ref{D1}) is exactly solvable 
and the solution yields asymptotically 
precise information about the cross-frequency
correlation, important for  analyzing
the  information transfer  and time reversal with
broadband signals in the channel 
described by the stochastic Schr\"odinger equation
with a $z$-white-noise potential
 \cite{pulsa} (see also \cite{BPZ}, \cite{DTF}, \cite{DTF2}).
 Moreover, eq. (\ref{fp})-(\ref{D1}) arises as the boundary layer equation for the
 two-frequency radiative transfer of {\em spherical wave}
 \cite{2f-rt-josa}.

The proof of Theorem~2 is  analogous to that
of Theorem~1  and for the sake of space we will not repeat
the argument. We refer the reader to  \cite{rad-arma}
for the needed minor modification.  More directly
eq. (\ref{fp}) can be obtained from
eq. (\ref{bolt}) in the limit $\theta\to 0$
by Taylor expanding the terms involving $\theta$ on the right hand side of (\ref{bolt}) up to second order in $\theta$. The 
first-order-in-$\theta$ terms are also first-order-in-$\bq$
and thus vanish due to the symmetry $K(\bp,\bq)
=K(\bp,-\bq)$. The remaining terms, after dividing by $\theta^2$ and passing to the limit,   become the right hand
side of (\ref{fp}).

For the proof of  Theorem~1 below, we  set $\theta=1$ for
ease of notation.
\section{Martingale formulation}

For tightness as well as identification of the limit,
 the following infinitesimal operator  $\cA^\ep$ will play an important role.
Let $V^\vas_z\equiv V(z/\ep^2,\cdot)$. Let $\mathcal{F}_z^\vas$ be the
$\sigma$-algebras generated by $\{V_s^\vas, \, s\leq z\}$  and
$\mathbb{E}_z^\vas$ the corresponding conditional expectation w.r.t. $\cF^\ep_z$.
Let $\cM^\ep$ be the space of measurable function adapted to $\{\cF^\ep_z, \forall t\}$ 
such that $\sup_{z<z_0}\IE|f(z)|<\infty$.
We say $f(\cdot)\in \cD(\cA^\ep)$, the domain of $\cA^\ep$, and $\cA^\ep f=g$
if $f,g\in \cM^\ep$ and for
$f^\delta(z)\equiv\delta^{-1}[\IE^\ep_z f(z+\delta)-f(z)]$
we have
\beqn
\sup_{z,\delta}\IE|f^\delta(z)|&<&\infty\\
\lim_{\delta\to 0}\IE|f^\delta(z)-g(z)|&=&0,\quad\forall
z.
\eeqn
Consider a special class of admissible functions
 $f(z)=\phi(\lan W_z^\vas, \Theta\ran),
 f'(z)=\phi'(\lan W_z^\vas, \Theta\ran),
 \forall \phi\in C^\infty(\IR)$
we have the following expression
from (\ref{weak}) and the chain rule 
\beq
\label{gen2}
 \cA^\vas
f(z)
&=&f'(z)\lt[\lan W^\ep_z,
{\bp}\cdot \nabla\Theta\ran + \frac{1}{\vas} \lan W^\ep_z,
\cv^*\Theta\ran\rt].
\eeq
In case of the test function $\Theta$ that is also a functional
of the media we have
\beq
\label{gen}
 \cA^\vas
 f(z)
 &=&f'(z)\lt[  \lan W^\ep_z,
 {\bp}\cdot \nabla\Theta\ran + \frac{1}{\vas} \lan W^\ep_z,
 \cv^*\Theta\ran +\lan W^\ep_z, \cA^\ep\Theta\ran\rt]
 \eeq
 and when $\Theta$ depends explicitly on the fast spatial variable
 \[
 \tilde{\bx}=\bx/\ep^{2\alpha}
 \]
 the gradient $\nabla$ is a sum of two terms: 
  \[
 \nabla=\nabla_\bx+\ep^{-2\alpha}\nabla_{\tilde{\bx}}
 \]
 where 
$\nabla_\bx$ is 
 the gradient w.r.t. the slow variable $\bx$  and $\nabla_{\tilde{\bx}}$ the gradient 
 w.r.t. the fast variable  $\xtil$. 

 A main property of $\cA^\ep$ is
that 
\beq
\label{12.2}
f(z)-\int^z_0 \cA^\ep f(s) ds\quad\hbox{is a  $\cF^\ep_z$-martingale},
\quad\forall f\in \cD(\cA^\ep).
\eeq
Also,
\beq
\label{mart}
\IE^\ep_s f(z)-f(s)=
\int^z_s \IE^\ep_s \cA^\ep f(\tau) d\tau\quad \forall s<z \quad\hbox{a.s.}
\eeq
(see \cite{Kur}).
We denote by $\cA$ the infinitesimal operator corresponding
to the unscaled process $V_z(\cdot)=V(z,\cdot)$.

\section{Proof of tightness}

In the sequel we will adopt the following notation 
\[
f(z)\equiv \phi(\lan  W^\ep_z,
\Theta\ran),\quad
 f'(z)\equiv \phi'(\lan  W^\ep_z,
\Theta\ran),\quad f''(z)\equiv
\phi''(\lan  W^\ep_z,
\Theta\ran),\quad 
\quad\forall \phi\in C^\infty(\IR).
\]
Namely, the prime stands for the differentiation w.r.t. the original argument (not $t$)
of $f, f'$ etc.

A family 
of processes $\{ W_z^\ep, 0<\ep<1\} $ in the Skorohod
space $ D([0,\infty);L^2_{w}(\IR^{2d})) 
$ is 
tight if and only if the family of 
processes
$\{\lan  W_z^\ep, \Theta\ran, 0<\ep <1\}
\subset D([0,\infty); L^2_{w}(\IR^{2d}))
$
is tight for all $\Theta\in C^\infty_c$ \cite{Fou}. 
We use the tightness  criterion of \cite{Ku}
(Chap. 3, Theorem 4), namely, 
we will prove:
Firstly,
\beq
\label{trunc}
\lim_{N\to \infty}\limsup_{\ep\to 0}\IP\{\sup_{z<z_0}|\lan  W_z^\ep, \Theta\ran|
\geq N\}=0,\quad\forall z_0<\infty.
\eeq
Secondly, for  each $\phi\in
C^\infty(\IR)$   there is a sequence
$f^\ep(z)\in\cD(\cA^\ep)$ such that for each $z_0<\infty$
$\{\cA^\ep f^\ep (z), 0<\ep<1,0<z<z_0\}$ 
is uniformly integrable and
\beq
\lim_{\ep\to 0} \IP\{\sup_{z<z_0} |f^\ep(z)-
\phi(\lan  W_z^\ep, \Theta\ran) |\geq
\delta\}=0,\quad \forall \delta>0.
\label{n59'}
\eeq
Then it follows that the laws of
$\{\lan  W_z^\ep, \Theta\ran, 0<\ep <1\}$ are tight in the space
of $D([0,\infty);\IR)$. 

To prove the tightness in the space $C([0,\infty); L^2_w(\IR^{2d}))$
let us  recall that $W^\ep_z\in C([0,\infty); L^2_w(\IR^{2d}))$ 
and  that the Skorohod metric and the uniform metric
induce the same topology on $C([0,\infty); L^2_w(\IR^{2d}))$.

Let us note first that condition (\ref{trunc}) is satisfied because the $L^2$-norm is
uniformly bounded.
The rest of the argument for tightness will be concerned with establishing
the second part of the criterion.

Consider now the expression
\beq
\label{n51}
{\tilde{\cL}^*_z\Theta(\bx,{\tilde{\bx}},\bp)}
&\equiv&
-{i}
\ep^{-2}\int_z^\infty\int  \Big[e^{i\bq\cdot\xtil/\sqrt{\ks_1}}\ks_1
\Theta(\bx,\bp-\frac{\bq}{2\sqrt{\ks_1}})-
e^{i\bq\cdot\xtil/\sqrt{\ks_2}}\ks_2\Theta(\bx,\bp+
\frac{\bq}{2\sqrt{\ks_2}})\Big]\\
&&\quad\quad\quad\quad\quad\times e^{i
(s-z)\bp\cdot\bq/\ep^{2\alpha}}
\IE_z^\ep\hat{V}^\ep_s(d\bq)
ds\nn
\eeq
or equivalently
\beq
\label{n50}
\cF^{-1}_2\cvtil\Theta(\bx,\xtil,\by)&=&
\ep^{-2}
\int_z^\infty
e^{-i\epal(s-z)\nabla_\by\cdot\nabla_{\tilde\bx}}
\lt[\IE^\ep_z\lt[\delta_{\ep} V^\ep_s\rt]
\cF^{-1}_2\Theta\rt](\bx,\xtil, \by) ds
\eeq
where $\delta_{\ep} V^\ep_s$ is defined by (\ref{diff}). 
It is straightforward to check that (\ref{n51}) solves the corrector equation
\beq
\label{corrector}
\lt[\ep^{-2\alpha}{\bp}\cdot\nabla_{\tilde\bx}+\cA^\ep\rt]
\tilde{\cL}^*_z\Theta=\ep^{-2}\cL^*_z\Theta
\eeq
Recall that $\grxtil$ and $\grx$ are the gradients
w.r.t. the fast variable $\xtil$ and the slow
variable $\bx$, respectively.

We have the fowllowing estimate.
\begin{lemma}
\label{Lemma1}
\beq
\nn
\lefteqn{\IE\lt[\cltil_z\Theta\rt]^2(\bx,\bp)}\\
& \leq&
\lt[\int^\infty_0\rho(s)ds \rt]^2
\int \Big|e^{i\bq\cdot\xtil/\sqrt{\ks_1}}\ks_1
\Theta(\bx,\bp+\frac{\bq}{2\sqrt{\ks_1}})-
e^{i\bq\cdot\xtil/\sqrt{\ks_2}}\ks_2\Theta(\bx,\bp-
\frac{\bq}{2\sqrt{\ks_2}})\Big|^2\Phi(\xi,\bq)d\xi d\bq.\nn
\eeq
\end{lemma}
\begin{proof}
Consider the following trial functions
in the definition of the maximal correlation coefficient
\beq
h&=&h_{s}(\bx,\bp)\nn\\
&=&{i}
\int \Big[e^{i\bq\cdot\xtil/\sqrt{\ks_1}}\ks_1
\Theta(\bx,\bp+\frac{\bq}{2\sqrt{\ks_1}})-
e^{i\bq\cdot\xtil/\sqrt{\ks_2}}\ks_2\Theta(\bx,\bp-
\frac{\bq}{2\sqrt{\ks_2}})\Big]
e^{i
k^{-1}(s-z)\bp\cdot\bq/\ep^{2\alpha}}
\IE_z^\ep\hat{V}^\ep_s(d\bq)\nn\\
g&=&g_{t}(\bx,\bp)\nn\\
&=&{i}
\int \Big[e^{i\bq\cdot\xtil/\sqrt{\ks_1}}\ks_1
\Theta(\bx,\bp+\frac{\bq}{2\sqrt{\ks_1}})-
e^{i\bq\cdot\xtil/\sqrt{\ks_2}}\ks_2\Theta(\bx,\bp-
\frac{\bq}{2\sqrt{\ks_2}})\Big]
e^{i
k^{-1}(t-z)\bp\cdot\bq/\ep^{2\alpha}}
\hat{V}^\ep_t(d\bq)\nn
\eeq
It is easy to see that 
$
h_s
\in  L^2(P,\Omega, \cF_{\ep^{-2}z})
g_t \in \in L^2(P,\Omega,
\cF^+_{\ep^{-2}t})
$
and their second moments are uniformly bounded
in $\bx,\bp,\ep$ since
\beqn
\IE[h_s^2](\bx,\bp)&\leq& \IE[g_s^2](\bx,\bp)\\
\IE[g_s^2](\bx,\bp) &=& \int
\Big|e^{i\bq\cdot\xtil/\sqrt{\ks_1}}\ks_1
\Theta(\bx,\bp+\frac{\bq}{2\sqrt{\ks_1}})-
e^{i\bq\cdot\xtil/\sqrt{\ks_2}}\ks_2\Theta(\bx,\bp-
\frac{\bq}{2\sqrt{\ks_2}})\Big|^2
\Phi(\xi,\bq) d\xi d\bq\label{n6}
\eeqn
which is uniformly bounded for any integrable 
spectral density
$\Phi$.

 From the definition
(\ref{correl}) we have
\beqn
\lt|\IE[h_s(\bx,\bp) h_t(\by,\bq)]\rt|&=
\lt|\IE\lt[h_s g_t\rt]\rt|
&\leq \rho(\ep^{-2}(t-z) )\IE^{1/2}\lt[h_s^2(\bx,\bp)\rt]
\IE^{1/2}\lt[g_t^2(\by,\bq)\rt].
\eeqn
Hence by  setting $s=t,\bx=\by,\bp=\bq$ first and the
Cauchy-Schwartz inequality we have
\beqn
\IE\lt[h_s^2\rt(\bx,\bp)]&\leq&\rho^2(\ep^{-2}(s-z))\IE[g_t^2(\bx,\bp)]\\
\lt|\IE\lt[ h_s(\bx,\bp) h_t(\by,\bq)\rt]\rt|
&\leq& \rho(\ep^{-2}(t-z) )\rho(\ep^{-2}(s-z))
\IE^{1/2}[g_t^2(\bx,\bp)]\IE^{1/2}[g_t^2(\by,\bq)],\quad
\forall s, t\geq z,\forall \bx, \by.
\eeqn

Hence
\beqn
\ep^{-4}\int^\infty_z\int^\infty_z \IE[h_s(\bx,\bp)
g_t(\bx,\bp)] ds dt
&\leq&\IE[g_t^2](\bx,\bp)
\lt[\int^\infty_0\rho(s)ds \rt]^2
\eeqn
which together with (\ref{n6}) yields
the lemma. 
\end{proof}

\begin{cor}
\label{cor1}
\beq
\label{n9}
\IE\lt[\bp\cdot\nabla_\bx\cltil_z\Theta\rt]^2(\bx,\bp)
& \leq&
\lt[\int^\infty_0\rho(s)ds \rt]^2
\int \Big|e^{i\bq\cdot\xtil/\sqrt{\ks_1}}\ks_1
\bp\cdot\nabla_\bx\Theta(\bx,\bp+\frac{\bq}{2\sqrt{\ks_1}})\\
&&-
e^{i\bq\cdot\xtil/\sqrt{\ks_2}}\ks_2\bp\cdot\nabla_\bx\Theta(\bx,\bp-
\frac{\bq}{2\sqrt{\ks_2}})\Big|^2
\Phi(\xi,\bq)d\xi d\bq.\nn
\eeq

\end{cor}
Inequality~(\ref{n9}) can be obtained 
from the expression
\beqn
{\bp\cdot\nabla_{\bx}\tilde{\cL}^*_z\Theta(\bx,{\tilde{\bx}},\bp)}
&\equiv&
{i}
\ep^{-2}\int_z^\infty\int e^{i\bq\cdot{\tilde{\bx}}}
\Big[e^{i\bq\cdot\xtil/\sqrt{\ks_2}}\ks_2
\bp\cdot\nabla_\bx\Theta(\bx,\bp+\frac{\bq}{2\sqrt{\ks_2}})\\
&&-
e^{i\bq\cdot\xtil/\sqrt{\ks_1}}\ks_1\bp\cdot\nabla_\bx\Theta(\bx,\bp-
\frac{\bq}{2\sqrt{\ks_1}})\Big]  e^{i
(s-z)\bp\cdot\bq/\ep^{2\alpha}}
\IE_z^\ep\hat{V}^\ep_s(d\bq)
ds
\nn
\eeqn
as in Lemma~\ref{Lemma1}.

We will need to estimate the iteration of $\cv$
and $\cvtil$: 
\beqn
\lefteqn{\cv^*\cltil_z\Theta(\bx,\xtil,
\bp)}\\
&=&-\frac{1}{\ep^{2}}\int^\infty_z ds
\int
\hat{V}^\ep_z(d\bq)
\IE_z^\ep[\hat{V}^\ep_s(d\bq')]
e^{i
(s-z)\bp\cdot\bq/\ep^{2\alpha}}\Big\{\Big[e^{i\bq\cdot\xtil/\sqrt{k_1}}e^{i\bq'\cdot\xtil/\sqrt{k_1}} \\
&&\times k_1^2\Theta(\bx,\bp+\frac{\bq'}{2\sqrt{k_1}}+
\frac{\bq}{2\sqrt{k_1}})-
e^{i\bq\cdot\xtil/\sqrt{k_1}}e^{i\bq'\cdot\xtil/\sqrt{k_2}} k_1 k_2\Theta(\bx,\bp-\frac{\bq'}{2\sqrt{k_2}}+\frac{\bq}{2\sqrt{k_1}})\Big] e^{i
(s-z)\bq'\cdot\bq/(2\ep^{2\alpha})}\label{n8}\\
&&-
\Big[e^{i\bq\xtil/\sqrt{k_2}}e^{i\bq'\cdot\xtil/\sqrt{k_1}}k_1k_2\Theta(\bx,\bp+\frac{\bq'}{2\sqrt{k_1}}-
\frac{\bq}{2\sqrt{k_2}})-e^{i\bq\xtil/\sqrt{k_2}}e^{i\bq'\cdot\xtil/\sqrt{k_2}}
k_2^2\Theta(\bx,\bp-\frac{\bq'}{2\sqrt{k_2}}-\frac{\bq}{ 2\sqrt{k_2}})\Big]\\
&&\times  e^{-i
(s-z)\bq'\cdot\bq/(2\ep^{2\alpha})}\Big\}
\nn\\
\lefteqn{\cltil_z\cltil_z\Theta(\bx,\xtil,
\bp)}\\
&=&-\frac{1}{\ep^{4}}\int^\infty_z \int^\infty_z ds dt
\int
\IE_z^\ep\hat{V}^\ep_s(d\bq)
\IE_z^\ep[\hat{V}^\ep_t(d\bq')]
e^{i
(s-z)\bp\cdot\bq/\ep^{2\alpha}}e^{i
(t-z)\bp\cdot\bq/\ep^{2\alpha}}\Big\{\Big[e^{i\bq\cdot\xtil/\sqrt{k_1}}e^{i\bq'\cdot\xtil/\sqrt{k_1}} \\
&&\times k_1^2\Theta(\bx,\bp+\frac{\bq'}{2\sqrt{k_1}}+
\frac{\bq}{2\sqrt{k_1}})-
e^{i\bq\cdot\xtil/\sqrt{k_1}}e^{i\bq'\cdot\xtil/\sqrt{k_2}} k_1 k_2\Theta(\bx,\bp-\frac{\bq'}{2\sqrt{k_2}}+\frac{\bq}{2\sqrt{k_1}})\Big] e^{i
(s-z)\bq'\cdot\bq/(2\ep^{2\alpha})}\label{n8'}\\
&&-
\Big[e^{i\bq\xtil/\sqrt{k_2}}e^{i\bq'\cdot\xtil/\sqrt{k_1}}k_1k_2\Theta(\bx,\bp+\frac{\bq'}{2\sqrt{k_1}}-
\frac{\bq}{2\sqrt{k_2}})-e^{i\bq\xtil/\sqrt{k_2}}e^{i\bq'\cdot\xtil/\sqrt{k_2}}
k_2^2\Theta(\bx,\bp-\frac{\bq'}{2\sqrt{k_2}}-\frac{\bq}{ 2\sqrt{k_2}})\Big]\\
&&\times  e^{-i
(s-z)\bq'\cdot\bq/(2\ep^{2\alpha})}
\Big\}
\nn
\eeqn
 which  can be more easily estimated by using
 (\ref{n50}) as follows. 
 First we have the expressions after the inverse Fourier
 transform
 \beq
\cF^{-1}_2\lt\{\cv^*\cltil_z\Theta\rt\}(\bx,\xtil,
\by)\label{n7'} &=&
\ep^{-2}\int^\infty_z
\delta_{\ep}
V^\ep_z e^{-i\epal
(s-z)\nabla_\by\cdot\nabla_{\tilde
\bx}}\lt[\IE_z[\delta_{\ep} V^\ep_s]\cF^{-1}_2\Theta\rt]
(\bx,\xtil, \by)ds\\
\label{prob-est}
\cF^{-1}_2\lt\{\cltil_z\cltil_z\Theta\rt\}(\bx,\xtil,
\by)\label{n7''} &=&
-\ep^{-4}\int^\infty_z
e^{-i\epal (t-z)\nabla_\by\cdot\nabla_{\tilde
\bx}}\lt\{\IE_z[\delta_{\ep}
V^\ep_t]e^{-i\epal (s-z)
\nabla_\by\cdot\nabla_{\tilde
\bx}}\rt.\\
&&\lt. \hspace{5cm} \lt[\IE_z[\delta_{\ep}
V^\ep_s]\cF^{-1}_2\Theta\rt]\rt\} (\bx,\xtil, \by)ds
dt.\nn
\eeq

 \begin{lemma}
\label{Lemma2}
\beqn
\IE\|\cv^*\cvtil\Theta\|_2^2
&\leq &C\lt(\int^\infty_0\rho(s)ds\rt)^2
 \IE[ V_z]^2
  \int \Big|e^{i\bq\cdot\xtil/\sqrt{\ks_1}}\ks_1
\Theta(\bx,\bp+\frac{\bq}{2\sqrt{\ks_1}})\\
&&-
e^{i\bq\cdot\xtil/\sqrt{\ks_2}}\ks_2\Theta(\bx,\bp-
\frac{\bq}{2\sqrt{\ks_2}})\Big|^2
 \Phi(\xi, \bq)d\xi d\bx d\bq d\bp
       \nn\\
\IE\|\cvtil\cvtil\Theta\|_2^2
&\leq &C\lt(\int^\infty_0\rho(s)ds\rt)^4
 \IE[ V_z]^2
  \int \Big|e^{i\bq\cdot\xtil/\sqrt{\ks_1}}\ks_1
\Theta(\bx,\bp+\frac{\bq}{2\sqrt{\ks_1}})\\
&&-
e^{i\bq\cdot\xtil/\sqrt{\ks_2}}\ks_2\Theta(\bx,\bp-
\frac{\bq}{2\sqrt{\ks_2}})\Big|^2
\Phi(\xi, \bq)d\xi d\bx d\bq d\bp
       \nn
       \eeqn
       for some constant $C$ independent of $\ep$.
\end{lemma}
\begin{proof}
Let us consider $\cvtil\cvtil\Theta$ here.
The calculation for  $\cv^*\cvtil\Theta$ is similar.

By the Parseval theorem and the unitarity of
$\exp{(i\tau \gry\cdot \grxtil)},\tau\in \IR,$
\beqn
\lefteqn{\IE\|\cltil_z\cltil_z\Theta\|_2^2=
\IE\|\cF^{-1}_2\cltil_z\cltil_z\Theta\|_2^2}\\
\commentout{
&=&\ep^{-8}\int\IE\lt\{\int^\infty_z
e^{-i\epal(t-z)\nabla_\by\cdot\nabla_{\tilde
\bx}}\lt\{\IE_z[\delta_\ep
V^\ep_t]e^{-i\epal (s-z)\nabla_\by\cdot\nabla_{\tilde
\bx}}\lt[\IE_z[\delta_\ep V^\ep_s]\cF^{-1}_2\Theta\rt]\rt\}
(\bx,\xtil, \by)ds dt\rt.\\
&&\lt.\int^\infty_z
e^{i\epal k^{-1}(t'-z)\nabla_\by\cdot\nabla_{\tilde
\bx}}\lt\{\IE_z[\delta_\ep
V^\ep_{t'}]e^{i\epal
k^{-1}(s'-z)\nabla_\by\cdot\nabla_{\tilde
\bx}}\lt[\IE_z[\delta_\ep V^\ep_{s'}]\cF^{-1}_2\Theta\rt]\rt\}
(\bx,\xtil,\by)ds'
dt'\rt\}d\bx
d\by\\&=&
\ep^{-8}\int\IE\lt\{\int^\infty_z
e^{-i\epal k^{-1}(t-t')/2\nabla_\by\cdot\nabla_{\tilde
\bx}}\lt\{\IE_z[\delta_\ep
V^\ep_t]e^{-i\epal k^{-1}(s-z)\nabla_\by\cdot\nabla_{\tilde
\bx}}\lt[\IE_z[\delta_\ep V^\ep_s]\cF^{-1}_2\Theta\rt]\rt\}
(\bx,\xtil,\by)ds dt\rt.\\
&&\lt.\int^\infty_z
e^{i\epal k^{-1}(t-t')/2\nabla_\by\cdot\nabla_{\tilde
\bx}}\lt\{\IE_z[\delta_\ep
V^\ep_{t'}]e^{i\epal
k^{-1}(s'-z)\nabla_\by\cdot\nabla_{\tilde
\bx}}\lt[\IE_z[\delta_\ep V^\ep_{s'}]\cF^{-1}_2\Theta\rt]\rt\}
(\bx,\xtil,\by)ds' dt'\rt\}d\bx d\by\\
&=&\ep^{-8}\int\IE\lt\{\int^\infty_z
\lt\{\IE_z[\delta_\ep
V^\ep_t]e^{-i\epal k^{-1}(s-z)\nabla_\by\cdot\nabla_{\tilde
\bx}}\lt[\IE_z[\delta_\ep V^\ep_s]\cF^{-1}_2\Theta\rt]\rt\}
(\bx,\xtil,\by)ds dt\rt.\\
&&\lt.\int^\infty_z
\lt\{\IE_z[\delta_\ep
V^\ep_{t'}]e^{i\epal
k^{-1}(s'-z)\nabla_\by\cdot\nabla_{\tilde
\bx}}\lt[\IE_z[\delta_\ep V^\ep_{s'}]\cF^{-1}_2\Theta\rt]\rt\}
(\bx,\xtil, \by)ds' dt'\rt\}d\bx d\by\\
}
&\leq&C_0\ep^{-8}\int\int^\infty_z
\lt|\IE\lt\{\IE_z[\delta_\ep
V^\ep_t]e^{-i\epal (s-z)\nabla_\by\cdot\nabla_{\tilde
\bx}}\lt[\IE_z[\delta_\ep V^\ep_s]\cF^{-1}_2\Theta\rt]\rt\}
(\bx,\xtil, \by)\rt|ds dt\\
&&\int^\infty_z
\lt|\IE\lt\{\IE_z[\delta_\ep
V^\ep_{t'}]e^{i\epal
(s'-z)\nabla_\by\cdot\nabla_{\tilde
\bx}}\lt[\IE_z[\delta_\ep V^\ep_{s'}]\cF^{-1}_2\Theta\rt]\rt\}
(\bx,\xtil,\by)\rt|ds' dt'd\bx d\by\\
&&+C_0\ep^{-8}\int\int^\infty_z
\lt|\IE\lt[\IE_z[\delta_\ep
V^\ep_t]\IE_z[\delta_\ep
V^\ep_{t'}]\rt]\rt| \lt|\IE\lt\{e^{-i\epal
(s-z)\nabla_\by\cdot\nabla_{\tilde
\bx}}\lt[\IE_z[\delta_\ep V^\ep_s]\cF^{-1}_2\Theta\rt]
(\bx,\by)\rt.\rt.\\
&&\lt.\lt.
\quad\quad \times e^{i\epal
(s'-z)\nabla_\by\cdot\nabla_{\tilde
\bx}}\lt[\IE_z[\delta_\ep V^\ep_{s'}]\cF^{-1}_2\Theta\rt]
(\bx,\xtil,\by)\rt\}\rt|ds dt ds' dt'd\bx d\by.
\eeqn
The last inequality follows from the Gaussian
property. Note that in the $\bx$ integrals above the
fast variable
$\xtil$ is integrated and is not treated 
as independent of $\bx$.

Let
\[
g(t)=\delta_\ep
V^\ep_t
\]
and
\[
h(s)=e^{-i\epal (s-z)\nabla_\by\cdot\nabla_{\tilde
\bx}}\lt[\delta_\ep V^\ep_s\cF^{-1}_2\Theta\rt].
\]
The same argument as that for Lemma~\ref{Lemma1} yields
\beqn
\lt|\IE[\IE_z[g(t)]\IE_z[ h(s)]]\rt|&\leq&
\IE^{1/2}[\IE_z[g(t)]^2]\IE^{1/2}[\IE_z[h(s)]^2]\\
&\leq&\rho(\ep^{-2}(t-z))\rho(\ep^{-2}(s-z))
\IE^{1/2}[g^2(t)]\IE^{1/2}[h^2(s)],\quad t, s\geq z;\\
\lt|\IE[\IE_z[g(t)]\IE_z[ g(t')]]\rt|&\leq&
\IE^{1/2}[\IE_z[g(t)]^2]\IE^{1/2}[\IE_z[g(t')]^2]\\
&\leq&\rho(\ep^{-2}(t-z))\rho(\ep^{-2}(t'-z))
\IE^{1/2}[g^2(t)]\IE^{1/2}[g^2(t')],\quad t, t'\geq z;\\
\lt|\IE[\IE_z[h(s)]\IE_z[ h(s')]]\rt|&\leq&
\IE^{1/2}[\IE_z[h(s)]^2]\IE^{1/2}[\IE_z[h(s')]^2]\\
&\leq&\rho(\ep^{-2}(s-z))\rho(\ep^{-2}(s'-z))
\IE^{1/2}[h^2(s)]\IE^{1/2}[h^2(s')],\quad s, s'\geq
z.
\eeqn

Combining the above estimates we get
\beqn
\IE\|\cltil_z\cltil_z\Theta\|_2^2
&\leq& 
C_1\lt(\int^\infty_0\rho(s)ds\rt)^4
\int \IE[\delta_\ep V^\ep_z]^2
\IE\lt[e^{-i\epal (s-z)\nabla_\by\cdot\nabla_{\tilde
\bx}}\lt[\delta_\ep V^\ep_s\cF^{-1}_2\Theta\rt]\rt]^2d\bx d\by\\
\commentout{
&\leq&C_2\lt(\int^\infty_0\rho(s)ds\rt)^4
\IE[ V^\ep_z]^2
\int\IE\lt[e^{-i\epal (s-z)\nabla_\by\cdot\nabla_{\tilde
\bx}}\lt[\delta_\ep V^\ep_s\cF^{-1}_2\Theta\rt]\rt]^2d\bx d\by\\
&\leq&C_2\lt(\int^\infty_0\rho(s)ds\rt)^4
\IE[ V^\ep_z]^2
\int \IE
 \lt\{e^{i\bq\cdot{\tilde{\bx}}}
 [\Theta(\bx,\bp+\bq/2)-
 \Theta(\bx,\bp-\bq/2)]\rt.\nn\\
 &&\lt. \hspace{6cm} \times
 e^{i
 (s-z)\bp\cdot\bq/\ep^{2\alpha}}
 \hat{V}^\ep_s(d\bq)\rt\}^2 d\bx d\bp\\
 }
 &\leq&C_2\lt(\int^\infty_0\rho(s)ds\rt)^4
 \IE[ V^\ep_z]^2
 \int 
 \Big|e^{i\bq\cdot\xtil/\sqrt{\ks_1}}\ks_1
\nabla_\bx\Theta(\bx,\bp+\frac{\bq}{2\sqrt{\ks_1}})\\
&&\quad\quad\quad-
e^{i\bq\cdot\xtil/\sqrt{\ks_2}}\ks_2\nabla_\bx\Theta(\bx,\bp-
\frac{\bq}{2\sqrt{\ks_2}})\Big|^2
 \Phi(\xi, \bq)d\xi d\bx d\bq d\bp
\eeqn

\end{proof}

Now let us consider the second moment of 
$\bp\cdot\nabla_\bx\cvtil\cvtil\Theta$
and $\cv^*\cvtil\cvtil\Theta$:
\beq
\lefteqn{\cF^{-1}_2\lt\{\pdgx\cltil_z\cltil_z
\Theta\rt\}(\bx,\xtil,\by)}\label{n10}\nn\\
&=&i
\ep^{-2}\nabla_\by\cdot\nabla_\bx \int^\infty_z
e^{-i\epal (t-z)\nabla_\by\cdot\nabla_{\tilde
\bx}}\lt\{\IE_z[\delta_\ep
V^\ep_t]e^{-i\epal (s-z)\nabla_\by\cdot\nabla_{\tilde
\bx}}\lt[\IE_z[\delta_\ep V^\ep_s]\cF^{-1}_2\Theta\rt]\rt\}
(\bx,\by)ds dt\nn\\
&=&i
\ep^{-2} \int^\infty_z
e^{-i\epal(t-z)\nabla_\by\cdot\nabla_{\tilde
\bx}}\lt\{\IE_z[\delta_\ep
V^\ep_t]e^{-i\epal (s-z)\nabla_\by\cdot\nabla_{\tilde
\bx}}\lt[\IE_z[\nabla_\by\delta_\ep
V^\ep_s]\cdot\cF^{-1}_2\nabla_\bx\Theta\rt]\rt\} (\bx,\by)ds
dt\nn\\ &&+
i\ep^{-2} \int^\infty_z
e^{-i\epal (t-z)\nabla_\by\cdot\nabla_{\tilde
\bx}}\lt\{\IE_z[\nabla_\by\delta_\ep
V^\ep_t]\cdot e^{-i\epal
(s-z)\nabla_\by\cdot\nabla_{\tilde
\bx}}\lt[\IE_z[\delta_\ep
V^\ep_s]\cF^{-1}_2\nabla_\bx\Theta\rt]\rt\} (\bx,\by)ds
dt\nn
\eeq

\beq
\lefteqn{\cF^{-1}_2\lt\{\cv^*\cltil_z\cltil_z\Theta\rt\}(\bx,\xtil,\by)}\label{n10'}\nn\\
&=&i
\ep^{-4} \delta_\ep V^\ep_z(\xtil,\by)\int^\infty_z
e^{-i\epal (t-z)\nabla_\by\cdot\nabla_{\tilde
\bx}}\lt\{\IE_z[\delta_\ep
V^\ep_t]e^{-i\epal (s-z)\nabla_\by\cdot\nabla_{\tilde
\bx}}\lt[\IE_z[\delta_\ep V^\ep_s]\cF^{-1}_2\Theta\rt]\rt\}
(\bx,\by)ds dt.\nn
\eeq
The same calculation as in Lemma~\ref{Lemma2} yields 
the following estimates:
\begin{cor}
\label{cor2}
\beqn
{\IE\|\bp\cdot\nabla_\bx\cltil_z\cltil_z\Theta\|_2^2}
&\leq&
C\lt(\int^\infty_0\rho(s)ds\rt)^4\Big\{
 \IE[\nabla_\by V^\ep_z]^2
 \int \Big|e^{i\bq\cdot\xtil/\sqrt{\ks_1}}\ks_1
\nabla_\bx\Theta(\bx,\bp+\frac{\bq}{2\sqrt{\ks_1}})\\
&&-
e^{i\bq\cdot\xtil/\sqrt{\ks_2}}\ks_2\nabla_\bx\Theta(\bx,\bp-
\frac{\bq}{2\sqrt{\ks_2}})\Big|^2
\Phi(\xi, \bq)d\xi d\bx
d\bq d\bp\\
&&+
 \IE[V^\ep_z]^2
 \int \Big|e^{i\bq\cdot\xtil/\sqrt{\ks_1}}\ks_1
\bp\cdot\nabla_\bx\Theta(\bx,\bp+\frac{\bq}{2\sqrt{\ks_1}})\\
&&-
e^{i\bq\cdot\xtil/\sqrt{\ks_2}}\ks_2\bp\cdot\nabla_\bx\Theta(\bx,\bp-
\frac{\bq}{2\sqrt{\ks_2}})\Big|^2
 \Phi(\xi,
\bq)d\xi d\bx d\bq d\bp\Big\};
\eeqn
\beqn
\IE\|\cv^*\cltil_z\cltil_z\Theta\|_2^2
&\leq&
C\lt(\int^\infty_0\rho(s)ds\rt)^4
 \IE[ V^\ep_z]^4
 \int \Big|e^{i\bq\cdot\xtil/\sqrt{\ks_1}}\ks_1
\Theta(\bx,\bp+\frac{\bq}{2\sqrt{\ks_1}})\\
&&-
e^{i\bq\cdot\xtil/\sqrt{\ks_2}}\ks_2\Theta(\bx,\bp-
\frac{\bq}{2\sqrt{\ks_2}})\Big|^2
 \Phi(\xi, \bq)d\xi d\bx d\bq d\bp
\eeqn
for some constant $C$ independent of $\ep$.
\end{cor}

Let
\begin{equation}
\label{1st}
f_1 (z)=
{\vas}
f'(z) \lan  W^\ep_z,\cvtil\Theta\ran
\end{equation}
be the 1-st perturbation of $f(z)$. 

\begin{prop}\label{prop:2}
\begin{enumerate}
$$\lim_{\ep\to 0}\sup_{z<z_0} \mathbb{E} |f_1(z)|=0,\quad
\lim_{\ep\to 0}\sup_{z<z_0} |f_1(z)|= 0
\quad \hbox{in probability}$$.
\end{enumerate}
\end{prop}

\begin{proof}
We have
\beq
\label{1.2}
\mathbb{E}[|f_1(z)|]\leq {\vas}\|f'\|_\infty
\| W_0\|_2 
\IE\|\tilde{\cV}^*_z\Theta\|_2
\eeq
and
\beq
\label{1.3}
\sup_{z< z_0} |f_1^\vas(z)|
 \leq {\vas}
\|f'\|_\infty  \| W_0\|_2
\sup_{z<z_0}\|\cvtil\Theta\|_2.
\eeq
Since $\cvtil\Theta$ is a Gaussian
process and $\cvtil\cvtil\Theta$ is  a $\chi^2$-process,
by an application of  Borell's 
inequality
\cite{Ad} we have
\beq
\label{G1}
\IE[\sup_{z<z_0}\|\cvtil\Theta\|^2_2]
&\leq&C \log{(\frac{1}{\ep})} 
\IE\|\cvtil\Theta\|^2_2;\\
\label{G2}
\IE[\sup_{z<z_0}\|\cvtil\cvtil\Theta\|^2_2] 
&\leq&C \log^2{(\frac{1}{\ep})} 
\IE\|\cvtil\cvtil\Theta\|^2_2,
\eeq
i.e. the supremum over $z<z_0$
inside the expectation can be over-estimated
by a 
$\log{(1/\ep)}$ factor for excursion on the
scale of any power of $1/\ep$. 
Hence the right side of (\ref{1.2}) is $O(\ep)$ while
the right side of (\ref{1.3}) is $o(1)$ in probability
by Chebyshev's inequality.
\end{proof}

Set $f^\ep(z)=f(z)+f_1(z)$. Then (\ref{n59'}) 
follows  immediately from Proposition~1.

Let us now prove  the uniform integrability of
$\cA^\ep f^\ep$. A straightforward calculation yields
\beqn
\cA^\vas f_1 &=&
{\ep} f'(z)\lan\wepz,\bp\cdot\nabla_\bx\cvtil \Theta\ran
+\ep f''(z)
\lan \wepz, \bp\cdot\nabla_\bx
\Theta\ran\lan\wepz, \cltil\Theta\ran\\
&&+f'(z)\lan
\wepz,\cv^*\cvtil\Theta\ran+
 f''(z)\lan \wepz, \cL_z^* \Theta\ran\lan \wepz, 
    \cvtil\Theta\ran -\frac{1}{\ep}f'(z)\lan \wepz,\cv^*\Theta\ran
\nn
\eeqn
and, hence
\beq
\label{25'}
\cA^\vas f^\ep(z)
&=&f'(z)\lan \wepz, \pdgx \Theta\ran
+f'(z)\lan \wepz, \cv^*\cvtil\Theta\ran
+f''(z)\lan\wepz,\cv^*\Theta\ran\lan\wepz,\cvtil\Theta\ran
\\
&&\quad
+{\ep}\lt[f'(z)\lan\wepz,\pdgx\cvtil\Theta\ran
+f''(z)\lan\wepz,\pdgx\Theta\ran\lan \wepz,\cvtil\Theta\ran\rt]\nn\\
&=&A_0(z)+A_1(z)+A_2(z)+R_1(z)
\nonumber
\eeq
where $A_1(z)$ and $A_2(z)$ are the $O(1)$ statistical coupling
terms.

\begin{prop}
\label{prop:3}
$$\lim_{\ep\to
0}\sup_{z<z_0}\IE|R_1(z)|^2=0$$.
\end{prop}
\begin{proof}
First we note that
 \beq
 |R_1|
  \nonumber
  &\leq& {\vas}
  \left[\|f''\|_\infty\| W_0\|^2_2
  \|\pdgx\Theta\|_2\|\cvtil\Theta\|_2+
  \|f'\|_\infty\|\psiep\|_2\|\pdgx(\cvtil\Theta)\|_2\rt].
 \label{1.10}
  \eeq
Clearly we have
$$\lim_{\ep\to
0}\sup_{z<z_0}\IE|R_1(z)|^2=0.$$
 by
Lemma~\ref{Lemma1}
  and Corollary~\ref{cor1}.  
\end{proof}

For the tightness criterion stated in the beginnings of the section,
it remains to show
\begin{prop}
\label{prop:1}
$\{\cA^\ep f^\ep\}$ are uniformly integrable.
\end{prop}

\begin{proof}
Let us show first that $\{A_i\}, i=0, 1,2,3$ are uniformly integrable.

For this we have the following estimates:
\bean
|A_0(z)|
&\leq&\|f'\|_\infty\| W_0\|_2\|\pdgx \Theta\|_2\\
|A_1(z)|
&\leq& \|f'\|_\infty\| W_0\|_2
\|\cv^*\cvtil\Theta\|_2\\
|A_2(z)|
&\leq& \|f''\|_\infty\| W_0\|^2_2
\|\cv^*\Theta\|_2\|\cvtil\Theta\|_2.
\eeqn
The second moments of the right hand side
of the above expressions are uniformly bounded as
$\ep\to 0$ by Lemmas~\ref{Lemma1} and \ref{Lemma2}
and hence $A_0(z), A_1(z), A_2(z)$ are uniformly
integrable. 
By Proposition~\ref{prop:3}, $R_1$
is uniformly integrable. Therefore $\cA^\ep f^\ep$ is
uniformly integrable by (\ref{25'}). 
  \end{proof}

\section{Identification of the limit}
The tightness just established permits
passing to the weak limit.
Our strategy for identifying the limit is to show directly
that in passing to the weak limit  the limiting process
solves the martingale problem with {\em null} quadratic variation.
This would imply the limiting equation is deterministic. 
The uniqueness
of solution to the  limiting deterministic equation for 
given   data then identifies the limit.

For this purpose,
we introduce the next perturbations $f_2, f_3$.
Let
\bea
\label{40.3}
A_2^{(1)}(\psi) &\equiv&\int\psi(\bx,\bp)
\cQ_1(\Theta\otimes\Theta)(\bx,\bp,\by,\bq)\psi(\by,\bq)\,d\bx d\bp\,d\by d\bq\\
A_1^{(1)}(\psi)&\equiv&\int\cQ'_1\Theta(\bx,\bp)\psi(\bx,\bp)\,\,d\bx d\bp,\quad
\forall \psi\in L^2(\IR^{2d})
\label{41.2}
\eea
where 
\beqn
    \cQ_1(\Theta\otimes\Theta)(\bx,\bp,\by,\bq)&=&
    \IE\lt[\cv^*\Theta(\bx,\bp)\cvtil\Theta(\by,\bq)\rt]
\eeqn
and
\[
\cQ'_1\Theta(\bx,\bp)=\IE\lt[\cv^*\cvtil\Theta (\bx,\bp)\rt].
\]
Clearly,
\beqn
A_2^{(1)}(\psi)=\mathbb{E}\lt[\lan\psi, \cv^*\Theta\ran
\lan\psi, \cvtil\Theta\ran\rt].
\eeqn

Let
\[
\cQ_2(\Theta\otimes\Theta)(\bx,\bp,\by,\bq)\equiv
\IE\lt[\cvtil\Theta(\bx,\bp)\cvtil\Theta(\by,\bq)\rt]
\]
and \[
\cQ'_2\Theta(\bx,\bp)=\IE\lt[\cvtil\cvtil\Theta (\bx,\bp)\rt].
\]
Let
\bean
A_2^{(2)}(\psi) &\equiv&\int\psi(\bx,\bp)
\cQ_2(\Theta\otimes\Theta)(\bx,\bp,\by,\bq)\psi(\by,\bq)\,\,d\bx d\bp\,d\by
d\bq\\
A_1^{(2)}(\psi)&\equiv&\int\cQ_2'\Theta(\bx,\bp)\psi(\bx,\bp)\,\,d\bx\,d\bp
\eean
Define
\beq
\label{2nd}
f_2(z) &=&\frac{\ep^2}{2}f''(z)
\lt[\lan\psiep,\cvtil\Theta\ran^2-A^{(2)}_2(\psiep)\rt]\\
f_3(z) &=&
\frac{\ep^2}{2}f'(z)\lt[\lan\psiep,\cvtil\cvtil\Theta\ran-
A^{(2)}_1(\psiep)\rt].
\label{3rd}
\eeq

\begin{prop}\label{prop:4}
$$ \lim_{\ep\to 0}\sup_{z<z_0} \mathbb{E}|f_2(z)|=0,\quad \lim_{\ep\to 0}\sup_{z<z_0}
\mathbb{E}|f_3(z)|=0. $$
\end{prop}
\begin{proof}
We have the bounds
\beqn
\sup_{z<z_0}\IE|f_2(z)|&\leq&
\sup_{z<z_0}{\vas^2}\|f''\|_\infty\lt[
\| W_0\|_2^2\IE\|\cvtil\Theta\|_2^2
+\IE[A^{(2)}_2(\psiep)]\rt]\\
\sup_{z<z_0}
\IE|f_3^\vas(z)|&\leq&
\sup_{z<z_0}
{\vas^2}\|f'\|_\infty
\lt[\| W_0\|_2\IE \|\cvtil\cvtil\Theta\|_2
+\IE[A_1^{(2)}(\psiep)]\rt].
\eeqn
A straightforward calculation shows that $\IE[A^{(2)}_2(\psiep)]$
and $\IE[A_1^{(2)}(\psiep)]$ stay uniformly bounded 
w.r.t. $\ep$. 
The right side of the above expressions then tends to zero as $\ep\to 0$
by Lemma~1 and 2.
\end{proof}

We have
\beqn
\cA^\vas f_2(z)&=&
f''(z)\left[-
\lan\tvas, \cv^*\Theta\ran\lan\tvas,\cvtil\Theta\ran + A^{(1)}_2(\tvas)\right]
+ R_2(z)\\
\cA^\vas f_3(z)&=&
f'(z)\left[-\lan
\tvas,\cv^*(\cvtil\Theta)\ran + A^{(1)}_1(\tvas)\right]+
R_3(z)
\eeqn
with
\beq
\nn
R_2(z)& =&\ep^2\frac{f'''(z)}{2}\left[
\lan\psiep,\pdgx\Theta\ran+\frac{1}{\ep}\lan\psiep,\cv^*\Theta\ran\rt]
\lt[\lan\psiep,\cvtil\Theta\ran^2-A_2^{(2)}(\psiep)\rt]\\
\nn
&& +
\ep^2f''(z)\lan\psiep,\cvtil\Theta\ran\lt[
\lan\psiep,\pdgx(\cvtil\Theta)\ran+\frac{1}{\ep}
\lan\psiep,\cv^*\cvtil\Theta\ran\rt]\\
&&-\ep^2f''(z)
\lt[\lan\psiep,\pdgx(G_\Theta^{(2)}\psiep)\ran+
\frac{1}{\ep}\lan\psiep,\cv^* G_\Theta^{(2)}\psiep\ran\rt]
\label{36}
\eeq
where 
$G_\Theta^{(2)}$ denotes the operator
\[
G_\Theta^{(2)}\psi\equiv \int \cQ_2(\Theta\otimes\Theta)(\bx,\bp,\by,\bq)\psi(\by,\bq)
\,d\by d\bq.
\]
Similarly
\beq
\label{n75}
R_3(z)&=&\ep^2
f'(z)\left[\lan\psiep,\pdgx(\cvtil\cvtil\Theta)\ran+\frac{\ks}{\ep}
\lan\psiep,\cv^*\cvtil\cvtil\Theta\ran\rt]\\
&&\nn +\frac{\ep^2}{2}
f''(z)\left[
\lan\psiep,\pdgx\Theta\ran+\frac{1}{\ep}\lan\psiep,\cv^*\Theta\ran\rt]
\lt[\lan\psiep,\cvtil\cvtil\Theta\ran-A_1^{(2)}(\psiep)\rt]\\
&&\nn-\ep^2f'(z)\lt[
\lan\psiep,\pdgx(\cQ_2'\Theta)\ran+\frac{1}{\ep}
\lan\psiep,\cv^*\cQ_2'\Theta\ran\rt].
\eeq

\begin{prop}\label{prop:5}
\[
\lim_{\ep\to 0}\sup_{z<z_0} \mathbb{E} |R_2(z)|=0,\quad \lim_{\ep \to 0}
\sup_{z<z_0} \mathbb{E}
|R_3(z)|=0.
\]
\end{prop}
\begin{proof}
Part of the argument is analogous to that  given for
Proposition~\ref{prop:4}.  The additional estimates that we need to consider
are  the following. 

In $R_2$: First we have 
\beqn
\nn
\lefteqn{\sup_{z<z_0}\ep^2
\IE\lt|\lan\psiep,\pdgx(G_\Theta^{(2)}\psiep)\ran\rt|}\\ 
&=&\ep^2\int \IE\lt[ \psiep(\bx,\bp)\psiep(\by,\bq)\rt]\IE\lt[
\bp\cdot\nabla_\bx
\cvtil\Theta(\bx,\bp)\cvtil\Theta(\by,\bq)\rt] 
d\bx d\by d\bp d\bq\\
&\leq& \ep^2\int \IE\lt[ \psiep(\bx,\bp)\psiep(\by,\bq)\rt]
\IE^{1/2}[\bp\cdot\nabla_\bx \cvtil\Theta]^2(\bx,\bp)\IE^{1/2}[
\cvtil \Theta]^2(\by,\bq) d\bx d\by d\bp d\bq
\commentout{********************
&=&
\ep^{-2}\IE\lt\{\int\int_z^\infty\int_z^\infty \int 
e^{i\bp'\cdot{\tilde{\bx}}}e^{i\bq'\cdot{\tilde{\by}}}
 e^{i
k^{-1}(s-z)\bp\cdot\bp'/\ep^{2\alpha}}
e^{i
k^{-1}(t-z)\bq\cdot\bq'/\ep^{2\alpha}} [\Theta(\by,\bq+\bq'/2)-
\Theta(\by,\bq-\bq'/2)]\nn\rt.\\
&&\lt.
\times\bp\cdot\nabla_\bx [\Theta(\bx,\bp+\bp'/2)-
\Theta(\bx,\bp-\bp'/2)]\IE\lt[ \psiep(\bx,\bp)\psiep(\by,\bq)\rt]
\IE_z^\ep[\hat{V}^\ep_s(d\bp')] \hat{V}^\ep_t(d\bq')
ds dt d\bx d\by d\bp d\bq\rt\}
********************}
\nn
\eeqn
which is $O(\ep^2)$ by using Lemma~\ref{Lemma1},
Corollary~\ref{cor1} and the fact  $ \IE\lt[
\psiep(\bx,\bp)\psiep(\by,\bq)\rt]\in L^2(\IR^{4d})$ 
in conjunction with the same
argument as in proof of Lemma~1; Secondly, we have 
\beqn
\sup_{z<z_0}\ep \IE\lt|\lan\psiep,\cv^* G_\Theta^{(2)}
\psiep\ran\rt|
&=&\sup_{z<z_0}\ep\|W_0\|_2 \IE\|\cv^*
\IE\lt[\cvtil\Theta\otimes\cvtil\Theta\rt]\psiep\|_2\nn\\
&=&\sup_{z<z_0}\ep\|W_0\|_2 \IE\|\invf\cv^*
\IE\lt[\invf\cvtil\Theta\otimes\invf\cvtil\Theta\rt]\invf\psiep\|_2.\nn
\commentout{***********************
&=&\sup_{z<z_0}\ep\|W_0\|_2
\IE\|\ep^{-4} \delta_\ep V^\ep_z(\bx,\by)\int^\infty_z
e^{-i\epal k^{-1}(t-z)\nabla_\by\cdot\nabla_{\tilde
\bx}}\IE\lt\{\IE_z[\delta_\ep
V^\ep_t]e^{-i\epal k^{-1}(s-z)\nabla_\by\cdot\nabla_{\tilde
\bx}}\lt[\IE_z[V^\ep_s]\cF^{-1}_2\Theta\rt]\rt\}
(\bx,\by)ds dt\|_2\nn\\
&\leq&\ep\|W_0\|_2 \IE\lt|\ep^{-4}\int^\infty_z\int^\infty_z 
\int
 e^{i\bp'\cdot\bx} e^{ik^{-1}(s-z)\bp\cdot\bp'\epal}
e^{\bp''\cdot\bx \epal}
\ep^{2\alpha-2}
 \lt[\Theta(\bx,\bp+\bp'/2+\ep^{2-2\alpha}\bp''/2)\rt.\rt.\\
&&\lt.\lt.
-\Theta(\bx,\bp-\bp'/2+\ep^{2-2\alpha}\bp''/2)
-\Theta(\bx,\bp+\bp'/2-\ep^{2-2\alpha}\bp''/2)
+\Theta
(\bx,\bp-\bp'/2-\ep^{2-2\alpha}\bp''/2)\rt]\hat{V}^\ep_z(d\bp'')
e^{i\bq'\cdot\by}\rt.\\
&&\lt. e^{ik^{-1}(t-z)\bq\cdot\bq'\epal} 
[\Theta(\by,\bq+\bq'/2)-\Theta(\by,\bq-\bq'/2)]\IE_z^\ep[\hat{V}^\ep_s(d\bp')]
\IE_z[\hat{V}^\ep_t(d\bq')] d\bx d\bp ds dt \psiep(\by,\bq) d\by
d\bq\nn
\rt|\\
&\leq& \ep\|W_0\|^2_2
\IE^{1/2}\lt\|\ep^{-4}\int^\infty_z\int^\infty_z 
\int
 e^{i\bp'\cdot\bx} e^{ik^{-1}(s-z)\bp\cdot\bp'\epal}
e^{\bp''\cdot\bx \epal}
\ep^{2\alpha-2}
 \lt[\Theta(\bx,\bp+\bp'/2+\ep^{2-2\alpha}\bp''/2)\rt.\rt.\\
&&\lt.\lt.
-\Theta(\bx,\bp-\bp'/2+\ep^{2-2\alpha}\bp''/2)
-\Theta(\bx,\bp+\bp'/2-\ep^{2-2\alpha}\bp''/2)
+\Theta
(\bx,\bp-\bp'/2-\ep^{2-2\alpha}\bp''/2)\rt]\hat{V}^\ep_z(d\bp'')
e^{i\bq'\cdot\by}\rt.\\
&&\lt. e^{ik^{-1}(t-z)\bq\cdot\bq'\epal} 
[\Theta(\by,\bq+\bq'/2)-\Theta(\by,\bq-\bq'/2)]
\IE\lt[\IE_z^\ep[\hat{V}^\ep_s(d\bp')]
\hat{V}^\ep_t(d\bq')\rt]
d\bx d\bp ds dt\rt\|_2^2 \nn
***********************}
\eeqn
Define
\beqn
h_s=e^{-i\epal(s-z)\nabla_\by\cdot\nabla_{\tilde\bx}} [\delta_\ep V^\ep_z \invf\Theta].
\eeqn
We then have
\beqn
\lefteqn{\IE\|\invf\cv^*
\IE\lt[\invf\cvtil\Theta\otimes\invf\cvtil\Theta\rt]\invf\psiep\|_2
}\\
&=&\IE\lt\{\int
\lt|\ep^{-4}\int \int^\infty_z \delta_\ep V^\ep_z(\bx,\by)
\IE\lt[\IE_z[ h_s(\bx,\by)]\IE_z[h_t(d\bx',d\by')]\rt]
\invf \wepz(\bx', \by')d\bx' d\by' ds dt\rt|^2 d\bx d\by\rt\}^{1/2}\\
\commentout{
&\leq&\IE^{1/2}\lt\{\int
\lt|\ep^{-4}\int^\infty_z |\delta_\ep V^\ep_z(\bx,\by)|
\rho(\ep^{-2}(s-z))\rho(\ep^{-2}(t-z))
\IE^{1/2}[h_s(\bx,\by)]^2\rt. \rt.\\
&&\lt.\lt.\quad
\int \IE^{1/2}[h_t(d\bx',d\by')]^2
|\invf \wepz(\bx', \by')|d\bx' d\by' ds dt\rt|^2 d\bx d\by\rt\}^{2}\\
}
&\leq&
\IE^{1/2}\lt\{\int
\lt|\ep^{-4}\int^\infty_z |\delta_\ep V^\ep_z(\bx,\by)|
\rho(\ep^{-2}(s-z))\rho(\ep^{-2}(t-z))
\IE^{1/2}[h_s(\bx,\by)]^2\rt. \rt.\\
&&\lt.\lt.\quad
\lt(\int \IE[h_t(d\bx',d\by')]^2 d\bx'd\by'\rt)
\lt(\int |\wepz(\bx', \bp')|^2d\bx' d\bp'\rt) ds dt\rt|^2 d\bx d\by\rt\}.
\eeqn
Recall that $\|W^\ep_z\|_2\leq \|W_0\|_2$ and
\beqn
\int \IE[h_t(d\bx',d\by')]^2 d\bx'd\by'
=\int
[\Theta(\bx,\bp+\bq/2)-
\Theta(\bx,\bp-\bq/2)]^2
\Phi(\xi,\bq) d\xi d\bq d\bx d\bp<\infty
\eeqn
so that
\beqn
\lefteqn{\IE\|\invf\cv^*
\IE\lt[\invf\cvtil\Theta\otimes\invf\cvtil\Theta\rt]\invf\psiep\|_2
}\\
\commentout{
&\leq& \|W_0\|_2 \IE^{1/2}\|h_s\|_2^2 \IE^{1/2}\lt\{\int
\lt|\ep^{-4}\int^\infty_z |\delta_\ep V^\ep_z(\bx,\by)|
\rho(\ep^{-2}(s-z))\rho(\ep^{-2}(t-z))
\IE^{1/2}[h_s(\bx,\by)]^2
ds dt\rt|^2 d\bx d\by\rt\}\\
}
&\leq& \|W_0\|_2 \IE^{1/2}\|h_s\|_2^2\lt( \sup_{\bx,\by}{
\IE[\delta_\ep V^\ep_z]^2}\rt)
\ep^{-8}\int^\infty_z 
\rho(\ep^{-2}(s-z))\rho(\ep^{-2}(t-z))
\\
&&\hspace{5cm}\times\rho(\ep^{-2}(s'-z))\rho(\ep^{-2}(t'-z))
\IE^{1/2}\|h_s\|_2^2
\IE^{1/2}\|h_{s'}\|_2^2
ds dt ds' dt' \\
&\leq& \|W_0\|_2 \IE^{3/2}\|h_s\|_2^2\lt( \sup_{\bx,\by}{
\IE[\delta_\ep V^\ep_z]^2}\rt)
\lt|\int^\infty_{0} 
\rho(s)
ds \rt|^2<\infty.
\eeqn
Recall from (\ref{n6}) that
\beqn
\IE\|h_s\|_2^2&=&\int
[\Theta(\bx,\bp+\bq/2)-
\Theta(\bx,\bp-\bq/2)]^2
\Phi(\xi,\bq) d\xi d\bq d\bx d\bp <\infty.\label{n16}
\eeqn
Hence 
\[
\sup_{z<z_0}\ep \IE\lt|\lan\psiep,\cv^* G_\Theta^{(2)}\wepz\ran\rt|=O(\ep). 
\]

In $R_3^\ep$:
\beqn
{\sup_{z<z_0}\ep 
\IE\lt|\lan\psiep,\cv^*\cvtil\cvtil\Theta\ran\rt|}
 &\leq& \ep
\|W_0\|_2\sup_{z<z_0}\IE\|\cv^*\cvtil\cvtil\Theta\|_2
\eeqn
which is $O(\ep)$ by Corollary~\ref{cor2}.

\commentout{*************************************
&\leq&\ep
\|W_0\|_2\sup_{z<z_0}\IE^{1/2}\lt\|\ep^{-4}\int_z^\infty\int^\infty_z
\int e^{i\bq\cdot\tilde{\bx}} \hat{V}_z(d\bp')
\IE_z^\ep\hat{V}^\ep_s(d\bq) \IE_z^\ep\hat{V}^\ep_t(d\bq')
e^{i\bq\cdot{\tilde{\bx}}}
e^{i\bq'\cdot{\tilde{\bx}}} 
e^{i
k^{-1}(s-z)\bp\cdot\bq/\ep^{2\alpha}}e^{i
k^{-1}(t-z)\bp\cdot\bq'/\ep^{2\alpha}}\rt.\nn\\
&&\lt.\ep^{2\alpha-2}\lt\{e^{i
k^{-1}(s-z)\bp'\cdot\bq\ep^{2-4\alpha}}e^{i
k^{-1}(t-z)\bp'\cdot\bq'\ep^{2-4\alpha}}\lt[[\Theta(\bx,\bp+\bp'/2+\bq'/2+\bq/2)-
\Theta(\bx,\bp+\bp'/2+\bq'/2-\bq/2)]\rt. \rt.\rt.\nn\\
&&\lt.\lt.\lt.\times e^{i
k^{-1}(s-z)\bq'\cdot\bq/(2\ep^{2\alpha})} -
[\Theta(\bx,\bp+\ep^{2-2\alpha}\bp'/2-\bq'/2+\bq/2)-
\Theta(\bx,\bp+\ep^{2-2\alpha}\bp'/2-\bq'/2-\bq/2)]\rt.\rt.\rt.\\
&&\lt.\lt.\lt.\times e^{-i
k^{-1}(s-z)\bq'\cdot\bq/(2\ep^{2\alpha})}\rt]-e^{-i
k^{-1}(s-z)\bp'\cdot\bq\ep^{2-4\alpha}}e^{-i
k^{-1}(t-z)\bp'\cdot\bq'\ep^{2-4\alpha}}\lt[[\Theta(\bx,\bp-
\ep^{2-2\alpha}\bp'/2+\bq'/2+\bq/2)\rt.\rt.\rt.\\
&&\lt.\lt.\lt.-
\Theta(\bx,\bp-\ep^{2-2\alpha}\bp'/2+\bq'/2-\bq/2)]  e^{i
k^{-1}(s-z)\bq'\cdot\bq/(2\ep^{2\alpha})} 
-[\Theta(\bx,\bp-\ep^{2-2\alpha}\bp'/2-\bq'/2+\bq/2)-\rt.\rt.\rt.\nn\\
&&\lt.\lt.\lt.-
\Theta(\bx,\bp-\ep^{2-2\alpha}\bp'/2-\bq'/2-\bq/2)] e^{-i
k^{-1}(s-z)\bq'\cdot\bq/(2\ep^{2\alpha})}\rt]\rt\}
ds dt\rt\|^2_2
\eeqn
which is $O(\ep)$ by using Assumption~3.
**********************************}
The other two terms in
$R_3$ have the respective expressions
\beqn
\ep^2
\IE\lt|\lan\psiep,\pdgx(\cQ_2'\Theta)\ran\rt|&\leq&
\ep^2\|W_0\|_2\IE^{1/2}\|\pdgx
\IE[\cvtil\cvtil\Theta]\|_2^2\\
&\leq& \ep^2\|W_0\|_2\|\IE[\pdgx\cvtil\cvtil\Theta]\|_2
\eeqn
which is $O(\ep^2)$ by Corollary~\ref{cor2}
and 
\beqn
{\ep}
\IE\lt|\lan\psiep,\cv^*\cQ_2'\Theta\ran\rt|&\leq&
\ep\|W_0\|_2\IE\|\cv^*\IE[\cvtil\cvtil\Theta]\|_2\\
&\leq&\ep\|W_0\|_2\lt(\sup_{\bx,\by}\IE^{1/2}\lt|\delta_\ep
V^\ep_z\rt|^2\rt) \IE^{1/2}\|\cvtil\cvtil\Theta\|_2^2
\eeqn
which is $O(\ep)$ by Lemma~\ref{Lemma2}.

\end{proof}

Consider the test function
$f^\ep(z)=f(z)+f_1(z)
+f_2(z)+f_3(z)$. We have
\beq
\label{2.67}
\lefteqn{\cA^\ep f^\ep(z)}\\
&=&
f'(z)\lan\psiep,\pdgx\Theta\ran+f''(z)
A_2^{(1)}(\psiep)+f'A_1^{(1)}(\psiep)
+R_1(z)+R_2(z)+R_3(z).\nn
\eeq
Set
\beq
\label{remainder}
R^\vas(z) = R_1(z) + R_2(z) +
R_3(z).
\eeq
It follows from Propositions 3 and 5 that
\[
\lim_{\ep \to 0}\sup_{z<z_0}\IE|R^\ep(z)|=0.
\]

\begin{prop}
\label{quad}
\[
\lim_{\ep\to 0}\sup_{z<z_0}\sup_{\|\psi\|_2=1}
A_2^{(1)}(\psi)
=0.
\]
\end{prop}
\begin{proof}
We have
\beqn
\nn
A_2^{(1)}(\psi) &=&\int\psi(\bx,\bp)
\cQ_1(\Theta\otimes\Theta)
(\bx,\bp,\by,\bq)\psi(\by,\bq)\,d\bx d\bp\,d\by d\bq\\
&=&\frac{1}{2}\int\psi(\bx,\bp)\widetilde\cQ_1(\bx,\bp,\by,\bq)
\psi(\by,\bq)\,d\bx d\bp\,d\by d\bq
\eeqn
where $\widetilde\cQ_1$ is defined by
\[
\widetilde\cQ_1(\bx,\bp,\by,\bq)=\lt[\cQ_1(\Theta\otimes\Theta)
(\by,\bq,\bx,\bp)+\cQ_1(\Theta\otimes\Theta)
(\bx,\bp,\by,\bq)
\rt].
\]
The symmetrized kernel has the following expressions
\beqn
\lefteqn{\widetilde\cQ_1(\bx,\bp,\by,\bq)}\\
&=&
\int^\infty_{-\infty}ds
\int \,d\bp' 
\check{\Phi}(s,\bp')
e^{i\bp'\cdot(\bx-\by)/\ep^{2\alpha}}e^{-is\bp\cdot\bp'\ep^{2-2\alpha}}
\Big[
e^{i\bp'\cdot\xtil/\sqrt{\ks_1}}\ks_1
\nabla_\bx\Theta(\bx,\bp+\frac{\bp'}{2\sqrt{\ks_1}})\\
&&-
e^{i\bp'\cdot\xtil/\sqrt{\ks_2}}\ks_2\nabla_\bx\Theta(\bx,\bp-
\frac{\bp'}{2\sqrt{\ks_2}})\Big]\Big[
e^{i\bp'\cdot\xtil/\sqrt{\ks_1}}\ks_1
\nabla_\bx\Theta(\bx,\bq+\frac{\bp'}{2\sqrt{\ks_1}})-
e^{i\bp'\cdot\xtil/\sqrt{\ks_2}}\ks_2\nabla_\bx\Theta(\bx,\bq-
\frac{\bp'}{2\sqrt{\ks_2}})\Big]\\
&=&
2\pi\int
e^{i\bp'\cdot(\bx-\by)/\ep^{2\alpha}}\Big[
e^{i\bp'\cdot\xtil/\sqrt{\ks_1}}\ks_1
\nabla_\bx\Theta(\bx,\bp+\frac{\bp'}{2\sqrt{\ks_1}})-
e^{i\bp'\cdot\xtil/\sqrt{\ks_2}}\ks_2\nabla_\bx\Theta(\bx,\bp-
\frac{\bp'}{2\sqrt{\ks_2}})\Big]\\
&&\times\Big[
e^{i\bp'\cdot\xtil/\sqrt{\ks_1}}\ks_1
\nabla_\bx\Theta(\bx,\bq+\frac{\bp'}{2\sqrt{\ks_1}})-
e^{i\bp'\cdot\xtil/\sqrt{\ks_2}}\ks_2\nabla_\bx\Theta(\bx,\bq-
\frac{\bp'}{2\sqrt{\ks_2}})\Big]
\Phi(\bp\cdot\bp'\ep^{2-2\alpha}, \bp')d\bp'.
\eeqn
which, as the inverse Fourier transform tends to zero
 uniformly outside any
neighborhood of
$\bx=\by$,  because of Assumption~1, and stays uniformly bounded everywhere. Therefore
the
$L^2$-norm of $\widetilde\cQ_1$ tends to zero
and the proposition follows.
\end{proof}

Similar calculation leads to the following
expression: For any real-valued, $L^2$-weakly
convergent sequence $\psi^\ep \to\psi$, we have
\beqn
{\lim_{\ep\to 0}A_1^{(1)}(\psi^\ep)}
&=&
\lim_{\ep\to 0} \int^\infty_0ds\int dw 
d\bq d\bx d\bp \,\,\psi^\ep(\bx,\bp)
\Phi(w,\bq) e^{isw}e^{-is \bp\cdot \bq\ep^{2-2\alpha}}\\
&&\times\Big\{ e^{-i\frac{s|\bq|^2}{2\sqrt{k_1}}\ep^{2-2\alpha}}
\Big[k_1k_2e^{i\bq\cdot\bx\ep^{-2\alpha}(\frac{1}{\sqrt{k_1}}
-\frac{1}{\sqrt{k_2}})}\Theta\Big(\bx,\bp+\frac{1}{2}
\big(\frac{1}{\sqrt{k_1}}+\frac{1}{\sqrt{k_2}}\big)\bq\Big)
-k_1^2\Theta(\bx, \bp) \Big]\\
&& +e^{i\frac{s|\bq|^2}{2\sqrt{k_2}}\ep^{2-2\alpha}}
\Big[k_1k_2e^{-i\bq\cdot\bx\ep^{-2\alpha}(\frac{1}{\sqrt{k_1}}
-\frac{1}{\sqrt{k_2}})}\Theta\Big(\bx,\bp-\frac{1}{2}
\big(\frac{1}{\sqrt{k_1}}+\frac{1}{\sqrt{k_2}}\big)\bq\Big)
-k_2^2\Theta(\bx, \bp) \Big]\Big\}\\
&=&k^2
\lim_{\ep\to 0} \int^\infty_0ds\int dw 
d\bq d\bx d\bp \,\,\psi^\ep(\bx,\bp)
\Phi(w,\bq) e^{isw}e^{-is\bp\cdot\bq \ep^{2-2\alpha}}\\
&&\times\Big\{ e^{-i\frac{s|\bq|^2}{2\sqrt{k}}\ep^{2-2\alpha}}
\Big[e^{i\bq\cdot\bx\beta/(2k^{1/2})}\Theta\Big(\bx,\bp+\frac{\bq}{\sqrt{k}}\Big)
-\Theta(\bx, \bp) \Big]\\
&&+ e^{i\frac{s|\bq|^2}{2\sqrt{k}}\ep^{2-2\alpha}}
\Big[e^{-i\bq\cdot\bx\beta/(2k^{1/2})}\Theta\Big(\bx,\bp-\frac{\bq}{\sqrt{k}}\Big)
-\Theta(\bx, \bp) \Big]\Big\}
\eeqn
where we have used (\ref{band}). 
Note that the integrand is invariant under the
change of variables:
$
s\to -s,\quad \bq\to -\bq.
$
Thus we can write 
\beqn
\lefteqn{\lim_{\ep\to 0}A_1^{(1)}(\psi^\ep)}\\
&=&
{k^2}\lim_{\ep\to 0} \int^\infty_{-\infty}ds\int dw 
d\bq d\bx d\bp \,\,\psi^\ep(\bx,\bp)
\Phi(w,\bq) e^{isw}e^{-is\bp\cdot\bq \ep^{2-2\alpha}} e^{-i\frac{s|\bq|^2}{2\sqrt{k}}\ep^{2-2\alpha}}\\
&&\times
\Big[e^{i\bq\cdot\bx\beta/(2k^{1/2})}\Theta\Big(\bx,\bp+\frac{\bq}{\sqrt{k}}\Big)
-\Theta(\bx, \bp) \Big]\\
&=&2\pi k^2\lim_{\ep\to 0} 
\int d\bq d\bx d\bp \,\,\psi^\ep(\bx,\bp)
\Phi\big(\ep^{2-2\alpha}(\bp+\frac{\bq}{2\sqrt{k}})\cdot\bq,\bq\big) 
\Big[e^{i\bq\cdot\bx\beta/(2k^{1/2})}\Theta\big(\bx,\bp+\frac{\bq}{\sqrt{k}}\big)
-\Theta(\bx, \bp) \Big]
\eeqn
from which we obtain 
\beqn
\bar A_1(\psi)&\equiv&\lefteqn{\lim_{\ep\to 0}A_1^{(1)}(\psi^\ep)}\\
&=&
\lt\{\begin{array}{ll}
2\pi  k^2
\int d\bq d\bx d\bp \,\,\psi(\bx,\bp)
\Phi(0,\bq) 
\Big[e^{i\bq\cdot\bx\beta/(2k^{1/2})}\Theta\big(\bx,\bp+\frac{\bq}{\sqrt{k}}\big)
-\Theta(\bx, \bp) \Big],\,\,\alpha\in (0,1)&\\
2\pi k^2
\int d\bq d\bx d\bp \,\,\psi(\bx,\bp)
\Phi\big((\bp+\frac{\bq}{2\sqrt{k}})\cdot\bq,\bq\big) 
\Big[e^{i\bq\cdot\bx\beta/(2k^{1/2})}\Theta\big(\bx,\bp+\frac{\bq}{\sqrt{k}}\big)
-\Theta(\bx, \bp) \Big],\,\,\alpha=1&.
\end{array}
\right.
\eeqn 

Recall that
\beq
\label{64}
M_z^\vas(\Theta)&=&
f(z)+f_1(z)+f_2(z)+f_3(z)
-
\int_0^zf'(z)\lan\tvas,\pdgx\Theta\ran\,ds\nn\\
&& - \int_0^z\left[f''(s)
A_2^{(1)}( W_s^\vas)+f'(s) A_1^{(1)}( W_s^\vas)\right]\,ds -\int_0^z
R^\vas(s)\,ds
\eeq
is a martingale.
The martingale property implies that for any finite
sequence
$ 0<z_1<z_2<z_3<...<z_n \leq z $,  $C^2$-function $f$ and 
bounded continuous  function $h$ with compact support,
we have 
\beq
\label{78}
\IE\lt\{h \lt(\lan W^\ep_{z_1}, \Theta\ran,
\lan W^\ep_{z_2},\Theta\ran,...,\lan W^\ep_{z_n},\Theta\ran\rt)
\lt[M^\ep_{z+s}(\Theta)-M^\ep_{z}(\Theta)\rt]\rt\}&=&0,\\
\quad \forall
s>0,\quad  z_1\leq z_2\leq\cdots\leq z_n\leq z.&&\nn
\eeq

Let
\[
\bar{\cA} f(z)\equiv
f'(z)\left[\lan
 W_z,\pdgx\Theta\ran+\bar{A}_1( W_z)\right].
\]
Here and below, by slight abuse of notation, $f(z)$ and
$f'(z)$ stand for $\phi(\lan W_z,\Theta\ran)$
and $\phi'(\lan W_z,\Theta\ran)$, respectively. 
In view of the results of
Propositions~\ref{prop:2},\ref{prop:3}, \ref{prop:1},
\ref{prop:4}, 
\ref{prop:5}, \ref{quad}  we see
that
\beqn
\IE\lt\{h \lt(\lan W^\ep_{z_1}, \Theta\ran,
\lan W^\ep_{z_2},\Theta\ran,...,\lan W^\ep_{z_n},\Theta\ran\rt)
\lt[f^\ep(z)-\phi(\lan W^\ep_z,\Theta\ran)\rt]\rt\}&=&0,\\
\IE\lt\{h \lt(\lan W^\ep_{z_1}, \Theta\ran,
\lan W^\ep_{z_2},\Theta\ran,...,\lan W^\ep_{z_n},\Theta\ran\rt)
\lt[\cA^\ep f^\ep(z)-\bar\cA\phi(\lan W^\ep_z,\Theta\ran)\rt]\rt\}&=&0.
\eeqn
With this and the tightness of $W^\ep_z$ 
we can pass to the limit
$\ep\to 0$ in  (\ref{78}), cf. \cite{EK}, Chapter 4, Theorem 8.10.  
Consequently that the limiting process satisfies the 
martingale property that
\beqn
\IE\lt\{h \lt(\lan W_{z_1}, \Theta\ran,
\lan W_{z_2},\Theta\ran,...,\lan W_{z_n},\Theta\ran\rt)
\lt[M_{z+s}(\Theta)-M_{z}(\Theta)\rt]\rt\}=0,\quad \forall s>0.
\eeqn
where
\begin{equation}
M_z(\Theta)=f(z)-\int_0^z \bar{\cA}f(s)
\,ds.
 \nn
\end{equation}
Then it follows that
\[
\IE\lt[ M_{z+s}(\Theta)-M_z(\Theta)|W_u, u\leq z\rt]=0,\quad
\forall z,s>0
\]
which proves that $M_z(\Theta)$ is a martingale.

Choosing $\phi(r) =r$ and $r^2$ 
we see that
\[
M_z^{(1)}(\Theta)=\lan  W_z,\Theta\ran -
\int_0^z \left[\lan  W_s,\pdgx\Theta\ran +
\bar{A}_1( W_s)\right]\,ds \]
is a martingale with the
null quadratic variation
\[ 
\left[M^{(1)}(\Theta),M^{(1)}(\Theta)\right]_z=
0.
\]
Thus
\[
f(z)-\int_0^z
\bigg\{f'(s)\left[\lan
 W_s,\pdgx\Theta\ran+\bar{A}_1( W_s)\right]
  \bigg\}\,ds= f(0),\quad\forall z>0.
  \]

Since $\lan\tvas,\Theta\ran$ is uniformly bounded
\[
\lt|\lan\tvas,\Theta\ran\rt|\leq \| W_0\|_2{\|\Theta\|}_2
\]
we have the convergence of the second moment
\[
\lim_{\ep\to 0}
\IE\left\{
{\lan\tvas,\Theta\ran}^2\right\}={\lan
 W_z,\Theta\ran}^2
 \]
 and hence the convergence in probability.

\end{document}